\documentclass[aps,prd,twocolumn,superscriptaddress,nofootinbib,showpacs,10pt]{revtex4-1}
\usepackage[utf8]{inputenc}
\usepackage[english]{babel}
\usepackage{amsmath}
\usepackage{amsfonts}
\usepackage{amssymb}
\usepackage{graphics}
\usepackage{graphicx}
\usepackage[left=1.5cm,right=1.5cm,top=1.5cm,bottom=1.5cm]{geometry}
\usepackage[usenames,dvipsnames,svgnames]{xcolor}  

\usepackage[linktocpage]{hyperref}   
\hypersetup{colorlinks=true,linkcolor=beamer@PRD, citecolor=beamer@PRD}
\definecolor{beamer@PRD}{RGB}{46,48,146}

\newcommand\myref[1]{\textcolor{beamer@PRD}{(}\ref{#1}\textcolor{beamer@PRD}{)}}

\begin{document}
\title{A squeezed review on coherent states and nonclassicality for non-Hermitian systems with minimal length}
\author{Sanjib Dey}\email{dey@iisermohali.ac.in}\email{sanjibdey4@gmail.com}
\affiliation{Department of Physics, Indian Institute of Science Education and Research Mohali, Sector 81, SAS Nagar, Manauli 140306, India}
\author{Andreas Fring}\email{a.fring@city.ac.uk}
\affiliation{Department of Mathematics, City, University of London, Northampton Square, London EC1V 0HB, UK}
\author{V\'eronique Hussin}\email{veronique.hussin@umontreal.ca}
\affiliation{Centre de Recherches Math\'ematiques $\&$ Department de Math\'ematiques et de Statistique, Universit\'e de Montr\'eal, Montr\'eal H3C 3J7, QC, Canada}
\begin{abstract}
It was at the dawn of the historical developments of quantum mechanics when Schr\"odinger, Kennard and Darwin proposed an interesting type of Gaussian wave packets, which do not spread out while evolving in time. Originally, these wave packets are the prototypes of the renowned discovery, which are familiar as ``coherent states" today. Coherent states are inevitable in the study of almost all areas of modern science, and the rate of progress of the subject is astonishing nowadays. Nonclassical states constitute one of the distinguished branches of coherent states having applications in various subjects including quantum information processing, quantum optics, quantum superselection principles and mathematical physics. On the other hand, the compelling advancements of non-Hermitian systems and related areas have been appealing, which became popular with the seminal paper by Bender and Boettcher in 1998. The subject of non-Hermitian Hamiltonian systems possessing real eigenvalues are exploding day by day and combining with almost all other subjects rapidly, in particular, in the areas of quantum optics, lasers and condensed matter systems, where one finds ample successful experiments for the proposed theory. For this reason, the study of coherent states for non-Hermitian systems have been very important. In this article, we review the recent developments of coherent and nonclassical states for such systems and discuss their applications and usefulness in different contexts of physics. In addition, since the systems considered here originate from the broader context of the study of minimal uncertainty relations, our review is also of interest to the mathematical physics community.
\end{abstract}


\maketitle
\tableofcontents

\section{\textbf{Introduction}} \label{sec1}
The terminologies \textit{coherent states}, \textit{squeezed states} and \textit{nonclassical states} have been so common in recent days that they can be found in almost every article in modern quantum optics and quantum information theory. Coherent states, which are, in a sense, the cornerstones of modern quantum optics, were proposed immediately after the birth of quantum mechanics by Schr\"odinger \cite{Schrodinger} followed by Kennard \cite{Kennard} and Darwin \cite{Darwin} in the form of a non-spreading Gaussian wave packets. The states were rediscovered later in 1963 by Glauber \cite{Glauber} in the form of eigenstates of the annihilation operator who expressed them in a fascinating form, which is familiar as coherent states today. Soon after its discovery, the subject has spread to almost all other areas of science very rapidly and the developments of the field and its applications have been breathtaking. 

Today, coherent states are found to exist in various branches of physics including mathematical physics \cite{Barut_Girardello,Perelomov, Gilmore,Gazeau_Klauder}, wavelets \cite{Ali_Antoine_Gazeau}, quantum gravity \cite{Bahr_Thiemann}, cosmology \cite{Hawking}, atomic and molecular physics \cite{Gazeau_Book}, to mention a few. Squeezed states as well as the other nonclassical states have originated mainly from the coherent states and, they are the source of entangled quantum states \cite{Kim_Son_Buzek_Knight} which are one of the fundamental ingredients for the study of quantum information theory. The span of the subject of coherent states and other related areas are almost limitless and, thus, it has not been possible to revise the whole subject in a single review article or in a book. Here we provide a partial list of references that have been devoted to several different aspects of coherent states \cite{Klauder_Skagerstam_Book,Perelomov_Book,Gerry_Knight_Book,Gazeau_Book, Ali_Antoine_Gazeau,Combescure_Robert_Book,Zhang_Review,Dodonov_Review, Dodonov_Manko_Book,Sanders_Review}.

Keeping aside the rapid progress of the subject of coherent states itself, in this article we intend to provide a concise review on the developments of coherent states for non-Hermitian systems primarily based on \cite{Dey_Fring_squeezed,Dey_Fring_Gouba_Castro,Dey, Dey_Hussin,Dey_Hussin_2, Dey2,Dey_Fring_Hussin}. Specifically, the systems that we study here originate form the noncommutative (NC) quantum mechanical structure with minimal length associated with the generalized uncertainty principle. The advancement of the coherent states for such systems are mainly based on the concepts of generalized coherent states \cite{Barut_Girardello,Stoler,Arik_Coon,Perelomov, Perelomov_Book,Manko_Marmo_Sudarshan_Zaccaria,Sivakumar} and, thus, their structure appears to be almost evident at first sight. However, we argue that they are highly nontrivial and challenging not only because of the complications of sophisticated mathematical structure of our system, but mostly, in the sense that they always emerge with the non-Hermitian structure, which is always challenging for the construction of concrete physical systems. Throughout this review, we not only discuss the construction of well-behaved coherent states for our systems, but also we shed light on the facts that these constructions may give rise to more degrees of freedom over the coherent states and nonclassical states of the standard harmonic oscillator, so that they may be utilized in  a more efficient way for further development of the subject area.
\section{\textbf{Formalism}}\label{sec2}
\subsection{Coherent states: general properties}
Coherent states are interesting superposition of infinitely many quantum states, whose dynamics most closely resemble the classical systems \cite{Fox_Choi,Dey_Fring_PRA}. It was Schr\"odinger \cite{Schrodinger}, who first discovered them in 1926 while searching for the solutions of the Schr\"odinger equation that satisfy the Bohr's correspondence principle. In his discovery, he constructed an interesting type of wave packet consisting of a large number of harmonic oscillator wave functions, which does not spread out with time and the behavior of which is very similar to a solitary wave as given by 
\begin{equation} \label{wavepacket}
\langle x \vert \alpha \rangle=\pi^{-1/4} \text{exp}\left(-\frac{1}{2}x^2+\sqrt{2}x \alpha-\frac{\alpha^2}{2}-\frac{\vert \alpha \vert ^2}{2}\right),
\end{equation}
with $\alpha$ being a complex parameter. Later in 1951, the same wave packet was derived in the opposite way by Iwata \cite{iwata}, who first considered the coherent states to be the eigenstates of the non-Hermitian annihilation operator
\begin{equation} \label{eigen}
\hat{a}\vert \alpha \rangle=\alpha \vert \alpha \rangle,
\end{equation}
and then successfully derived the same wavepacket \myref{wavepacket}. Afterwards, many authors obtained the same equation \myref{wavepacket} from many different arguments and finally it was Glauber \cite{Glauber}, who actually carried out a more systematic analysis to express it more compactly in terms of the Fock basis
\begin{equation} \label{GlauberCoherent}
\vert \alpha \rangle=e^{-\vert \alpha \vert^2/2}\displaystyle\sum_{n=0}^\infty\frac{\alpha^n}{\sqrt{n!}}\vert n \rangle, \qquad \forall\quad \alpha \in \mathbb{C},
\end{equation}
and entitled them as the ``coherent states" for the first time in the literature. It is, by now, well-known that the coherent states can be constructed from any of the following definitions: (i) as eigenstates of the annihilation operator $a\vert\alpha\rangle=\alpha\vert\alpha\rangle$, (ii) by applying the Glauber's unitary displacement operator, $D(\alpha)=\text{exp}(\alpha a^\dagger-\alpha^\ast a)$ on the vacuum state and (iii) as quantum states that minimize the uncertainty relation $\Delta x^2\Delta p^2=\hbar^2/4$, with equal uncertainties in each coordinate, $\Delta x^2=\Delta p^2$ \cite{Loudon_Knight}. However, every coherent state does not satisfy all of the above properties at a same time; for instance, see \cite{Antoine_Gazeau_Monceau_Klauder_Penson,Dey_Fring_Gouba_Castro}, where the first two properties have been satisfied, but not the third one. Coherent states that satisfy all of the three properties, are usually specified to \textit{intelligent states} \cite{Aragone_Guerri_Salamo_Tani,Trifonov5,Trifonov6,Dodonov_Review} or sometimes \textit{squeezed coherent states}. Glauber coherent state \myref{GlauberCoherent} is an ideal example of intelligent state.

Coherent states have several interesting mathematical features that are very different from the usual Fock states. For instance, two different coherent states are not orthogonal to each other due to the property that the annihilation operator is not a self adjoint operator by definition. There are many other interesting features of coherent states, which are well documented in the literature and, therefore, we are not going to discuss them in detail. Rather, we refer the readers to \cite{Klauder_Skagerstam_Book,Perelomov_Book,Gerry_Knight_Book,Gazeau_Book, Ali_Antoine_Gazeau,Combescure_Robert_Book,Zhang_Review,Dodonov_Review, Dodonov_Manko_Book,Sanders_Review}.
\subsection{Nonclassicality}
The clue to the explication of nonclassicality lies within its name, it simply means that any state which is not classical is nonclassical. However, the concept of nonclassicality originally follows from Glauber and Sudarshan's convention \cite{Glauber,Glauber3,Sudarshan2}. According to them, any state which is less classical than the coherent state is nonclassical. A more precise definition emerges from the Glauber-Sudarshan's $P$-function, which was introduced in \cite{Glauber3} to represent thermal states and in \cite{Sudarshan2} for arbitrary density matrices
\begin{eqnarray}
&& \rho=\int P(\alpha)|\alpha\rangle\langle \alpha| d\text{Re}\alpha~d\text{Im}\alpha, \\
&& \text{with} ~\int P(\alpha) d\text{Re}\alpha~d\text{Im}\alpha=1. \notag
\end{eqnarray}
For coherent states, the weight function $P(\alpha)$ can be interpreted as a probability density, as in this case the $P$-function is a delta function. Glauber defined the nonclassical states as the states for which the $P$-distribution fails to be a probability density. More specifically, if the singularities of $P$-functions are either of types stronger than those of the delta functions (e.g. derivatives of delta function) or they are negative, the corresponding states have no classical analogue. There are few other related techniques by which one can define the nonclassicality of a state, for instance by, Husimi $Q$-representations \cite{Husimi}, Wigner representations \cite{Wigner}, which can be found in any standard textbook of quantum optics \cite{Mandel_Wolf_Book,Scully_Zubairy_Book,Gerry_Knight_Book, Klauder_Sudarshan_Book,Walls_Milburn_Book,Agarwal_Book}. A completely different statement was given in \cite{Johansen}, where the author argued that a quantum state may be nonclassical even though the $P$-distribution is a probability density. Rather, the nonclassicality is associated with the failure of the Margenau-Hill distribution to be a probability distribution. Nevertheless, we will stick to the convention of Glauber and Sudarshan throughout our discussion. Although, the notion of nonclassicality was established by Glauber, the name nonclassical states, nonclassical effects, nonclassical light appeared much later in the literature; see, for instance, \cite{Helstrom,Lugiato}.

Nonclassicality can also be explained from a slightly different angle. For instance, one can consider a Gaussian wave packet with unequal variances of two quadratures, whose $P$-function in the special case of the statistically uncorrelated quadrature components reads as \cite{Mollow_Glauber}
\begin{equation}\label{PG}
P_G(\alpha)=\mathcal{N}\left[-\frac{(\text{Re}\alpha-a)^2}{\sigma_x-1/2}-\frac{(\text{Im}\alpha-b)^2}{\sigma_p-1/2}\right],
\end{equation}
where $a$ and $b$ are the position of the center of the distribution in the $\alpha$ plane and $\mathcal{N}$ stands for the normalization constant. The function \myref{PG} exists as a normalizable distribution only for $\sigma_x\geq 1/2$ and $\sigma_p\geq 1/2$. Therefore, the state with the variances of one of the quadratures being less than $1/2$ does not correspond to a Gaussian distribution and, therefore, the corresponding state is nonclassical. This statement is not only true for Gaussian states, but holds for any state. Indeed, it is possible to express the quadrature variance in terms of the $P$-function as follows
\begin{equation}
\sigma_x=\frac{1}{2}\int P(\alpha)\left[\left(\alpha+\alpha^\ast-\langle\hat{a}+\hat{a}^\dagger\rangle\right)^2+1\right]d\text{Re}\alpha~d\text{Im}\alpha.
\end{equation}
Now, if $\sigma_x<1/2$, the $P$-function must be negative and, therefore, the state can not be interpreted as classical \cite{Mollow_Glauber}. This is one of the strong evidences of nonclassicality and the phenomena is usually known as \textit{quadrature squeezing}, which we will discuss with more details in the next subsection.

As a matter of fact, all pure states except the coherent states are nonclassical with respect to the their physical properties \cite{Aharonov_Falkoff} as well as the formal definitions with respect to the $P$-function given above \cite{Hillery2}. Nonclassicality of the Fock states and their finite superpositions was mentioned in \cite{Titulaer_Glauber}. However, speaking of nonclassical states, people usually do not have any arbitrary pure quantum states in mind, but the states with more or less useful and distinctive properties. For instance, superposition of two coherent states, which are often known as the Schr\"odinger cat states, can be used as qubit in quantum information processing \cite{Gao}. There are several other well-known nonclassical states; such as, squeezed states \cite{Walls, Loudon_Knight}, photon-added coherent states (PACS) \cite{Agarwal_Tara}, pair coherent states \cite{Xia_Guo}, photon subtracted squeezed states \cite{Wakui}, etc. For more information on nonclassical states one may refer to, for instance, the review article \cite{Dodonov_Review}. We will discuss some of them for our non-Hermitian systems in this article.
\subsection{Coherence versus nonclassicality: Methods of identification}
Given the definitions of nonclassical states that we have discussed in the last section, it is fairly straightforward to identify them. However, sometimes it becomes important to not just test a binary notion of classical versus nonclassical but to develop a concept that quantifies the amount of nonclassicality it possesses \cite{Lee,Marian}. For instance, higher nonclassical states are more useful in quantum information theory as well as they reduce optical noise in one quadrature component with respect to that of the coherent states. In order to study such type of properties, in this section, we discuss some techniques that may be utilized for this purpose.
\subsubsection{Quadrature squeezing}
In quantum optics, quadratures are usually referred to the dimensionless position and momentum operators
\begin{equation}\label{quadrature}
y=\frac{1}{2}(a+a^\dagger), \quad z=\frac{1}{2i}(a-a^\dagger).
\end{equation}
Any quadrature is said to be squeezed if the variance of that quadrature becomes less than the right hand side of the uncertainty relation, which is $1/2$ for the systems satisfying the Heisenberg's uncertainty relation, or for a general systems which obeys the Robertson-Schr\"odinger uncertainty relation, it is $\frac{1}{2}\big\vert\langle[X,P]\rangle\big\vert$. Here, $X$ and $P$ are the position and momentum operators of the models obeying the generalized Robertson-Schr\"odinger uncertainty relation. The above situation (squeezing in one quadrature) can occur in two scenarios (by considering a system satisfying Heisenberg's uncertainty relation), when the uncertainty relation is saturated $\Delta x\Delta p=1/2$, as well as, in the usual case when $\Delta x\Delta p>1/2$. The latter case is more usual and happens more frequently, whereas, the former case is more interesting, since in this case the uncertainty relation is minimized with one of its quadratures being squeezed. This phenomenon is very rare in the literature, and it corresponds to an excellent type of nonclassical states, which is sometimes referred as \textit{ideal squeezed states} \cite{Caves,Milburn_Walls5,Trifonov7,Gerry_Knight_Book}.
\subsubsection{Photon number squeezing}
The meaning of photon number squeezing is that the photon number distribution of the state is narrower than the average number of photons $\langle (\Delta n)^2\rangle < \langle n\rangle$ , with $n=a^\dagger a$ being the number operator. A straightforward way to identify the photon statistics of any states as indicated by Mandel \cite{Mandel} is to calculate the Mandel parameter $Q$:
\begin{equation}\label{Mandel}
Q=\frac{\langle (\Delta n)^2\rangle}{\langle n\rangle}-1.
\end{equation}
For states with $Q=0$, the statistics are Poissonian, while $Q>0$ and $Q<0$ correspond to the cases of super-Poissonian and sub-Poissonian statistics, respectively. For a coherent state, it is well known that the photon distribution is Poissonian $(Q=0)$ with a mean of $\langle n\rangle$. While, the nonclassicality corresponds to the case of sub-Poissonian statistics $Q<0$. An alternative approach to realize the existence of photon number squeezing is to study the second order correlation function (for zero delay time) \cite{Glauber2}
\begin{equation}
g^{(2)}(0)=\frac{\left\langle(a^\dagger)^2a^2\right\rangle}{\langle a^\dagger a\rangle^2}.
\end{equation}
The number squeezing applies to the scenario when $g^{(2)}(0)<1$, physically which is referred to as a light field with photons more equally spaced than a coherent laser field and the phenomenon is popularly known as \textit{photon antibunching}.
\subsubsection{Photon distribution function}
A simple study of photon distribution function of a state, say $|\alpha\rangle$
\begin{equation}\label{PhotonDistribution}
P_n:=\big\vert\langle n|\alpha\rangle\big\vert^2,
\end{equation}
with $n$ being the number operator, can indicate the nonclassical behavior of the state. An oscillatory behavior of the function \myref{PhotonDistribution} corresponds to a nonclassical system, whereas, for classical-like states the distribution function \myref{PhotonDistribution} is of Gaussian type.
\subsubsection{Revival structure}
Further insights into the nonclassical behavior of a time-dependent system can be obtained from the study of the revival structure of the wave packets. For a general wave packet of the form $\psi=\sum c_n\phi_n$, with the mean of the distribution being $n=\bar{n}$ and energy $E_{\bar{n}}$, the quantum revival \cite{Eberly} is a periodic recurrence of the quantum wave function from its original form during the time evolution. A partial revival for which the initial form of the wavefunction is not recovered completely is known as \textit{fractional revival}. If the quantum revival or regaining happens exactly at its classical periods $T_{\text{cl}}=2\pi\hbar/|E'_{\bar{n}}|$, the structure correspond to a classical-like system. Whereas, if it occurs more than once within the classical period periodically at times $T_{\text{rev}}=4\pi\hbar/|E''_{\bar{n}}|$, it was argued in \cite{Averbukh} that the phenomenon corresponds to a system where the photon statistics is sub-Poissonian and, thus, the corresponding state is nonclassical in nature. In case of the existence of a revival structure, fractional revivals may also occur at times $p/q T_{\text{rev}}$, with co-prime integers $p,q$. If, the quantum revival occurs only once within the classical period, it is usually said that the system possess a \textit{revival} structure, while if it happens twice within the same period, it owns a \textit{super-revival} structure and in case of thrice a \textit{super-super-revival} structure, and so on. Systems with super-revival structure are more nonclassical than those of having revival structure. The easiest way to study the revival or super-revival structure is to compute the auto-correlation function of a state $|\gamma,\phi\rangle$
\begin{equation}\label{AutoCorrelation}
A(t):=\big\vert\langle\gamma,\phi|\gamma+\omega t,\phi\rangle\big\vert^2,
\end{equation}
where $\gamma$ is a parameter of the system and $\phi$ labels the state.
\subsubsection{Beam splitter entanglement}\label{SubSec25}
A beam splitter is a familiar optical interferometer, which has two input and two output ports. The lights passing through the input ports are partly reflected and partly transmitted with the amplitude reflection and transmission coefficients being $r$ and $t$, respectively. The quantum version of the classical beam splitter is obtained by replacing the incoming electromagnetic fields with a set of annihilation operators $a$ and $b$ corresponding to two different inputs \cite{Gerry_Knight_Book}. The output fields are, then, realized with the unitarily transformed operators $c=\mathcal{B}a \mathcal{B}^\dagger$ and $d=\mathcal{B}b \mathcal{B}^\dagger$, such that 
\begin{equation}\label{OutputComm}
[c,c^\dagger]=1 \qquad \text{and} \qquad [d,d^\dagger]=1.
\end{equation}
The unitary operator $\mathcal{B}$ is known as the beam splitter operator
\begin{equation}
\mathcal{B}=e^{\frac{\theta}{2}(a^\dagger b e^{i\phi}-a b^\dagger e^{-i\phi})}~,
\end{equation}
where $\theta$ denotes the angle of the beam splitter and $\phi$ is the phase difference between the reflected and transmitted fields. The conditions \myref{OutputComm} impose the restriction on the reflection and transmission amplitudes, $\vert r\vert^2+\vert t\vert^2=1$, with $r=-e^{-i\phi}\sin(\theta/2)$ and $t=\cos(\theta/2)$. For a $50:50$ beam splitter, $r$ and $t$ are naturally equal in amplitude, $\vert r\vert=\vert t\vert=1/\sqrt{2}$. The effect of the beam splitter operator on a bipartite input state composed of a usual Fock state $\vert n\rangle$ at one of the inputs and a vacuum state $\vert 0\rangle$ at the other, is well-known  \cite{Kim_Son_Buzek_Knight}
\begin{equation}\label{BeamFock}
\mathcal{B}\vert n\rangle_a \vert 0\rangle_b=\displaystyle\sum_{q=0}^n \begin{pmatrix}
n \\ q
\end{pmatrix}^{1/2}t^q r^{n-q}~\vert q\rangle_c\vert n-q\rangle_d~.
\end{equation}
The output of the beam splitter \myref{BeamFock} can be used in a reduced density matrix to calculate any type of entanglement entropy; such as, linear entropy, von-Neumann entropy, etc. One of the exciting features of a quantum beam splitter is that it produces entangled output states, if at least one of the input fields is nonclassical \cite{Kim_Son_Buzek_Knight,Xiang-Bin}. That is why one does not obtain the entangled states in the output ports, when one transmits coherent states through the input ports \cite{Kim_Son_Buzek_Knight}. One of the biggest advantages of using beam splitter as a test of nonclassicality includes the comparison of the entanglement entropy among many states and, thus, it may help us to recognize the nonclassical state that possesses the highest degree of nonclassicality among many states.
\section{\textbf{Non-Hermtian systems in minimal length scenario}}\label{sec3}
Having introduced the general features of coherent states and nonclassicality, in this section, let us discuss a system on which our article is based on, i.e. some non-Hermitian models based on quantum mechanics in NC space and, then, we discuss how these models can be applied to physical systems like coherent states.   
\subsection{Noncommutative quantum mechanics}\label{SubSec31}
The original proposal of space-time noncommutativity is very old and was introduced in the pioneering days of quantum field theory most notably by Heisenberg, who argued that one could use a NC structure of space-time at very small length scales to introduce the effective ultraviolet cut-off to regularize the ultraviolet divergence. The idea was given a proper mathematical structure for the first time by Snyder in 1947 \cite{Snyder}. Immediately after this, Yang extended Snyder's idea by replacing the algebra of noncommuting linear operators by the algebra of functions to describe a general geometrical structure \cite{Yang}. However, all of these suggestions were  ignored at that time, perhaps mainly due to the failure of making accurate experimental predictions of the theory, but mostly because of its timing. At around the same time, the renormalization group program of quantum field theory finally was becoming successful at accurately predicting numerical values for physical observables in quantum electrodynamics and, therefore, the theory of NC space-time went through a long period of ostracism.

However, the theory was reborn with a very simple and elegant Lorentz-covariant version introduced by Seiberg and Witten \cite{Seiberg_Witten}, who showed that the string theory can be realized as an effective quantum field theory in a NC space-time at a certain low-energy limit. Some important mathematical developments of the 1980s have also contributed to this rebirth, for instance, Connes \cite{Connes} and Woronowicz \cite{Woronowicz} revived the notion by introducing a differential structure in the NC framework and, NC theories have been an area of intense research since then. For further details on the subject, we refer the readers to some reviews, for instance \cite{Garay,Connes_Book,Madore_Book,Douglas_Nekrasov,Szabo}.

Nevertheless, the theory of noncommutativity has evolved from time to time and has shown its usefulness in different areas of modern physics \cite{Doplicher,Aschieri, Castro_Kullock_Toppan,Dey_Fring_Time, Dey_Fring_Mathanaranjan,Gouba_Review}. Besides, some well-known versions of it, some natural and desirable possibilities may arise when the canonical space-time commutation relation is deformed by allowing general dependence of position and momentum \cite{Kempf_Mangano_Mann,Das_Vagenas,Gomes_Kupriyanov,Bagchi_Fring, Quesne_Tkachuk}. In such scenarios, the Heisenberg uncertainty relation necessarily modifies to a generalized version to the so-called \textit{generalized uncertainty principle} (GUP). Over last two decades, it is known that within this framework, in particular, where the space-time commutation relation involves higher powers of momenta, explicitly
leads to the existence of nonzero minimal uncertainty in position coordinate, which is familiar as the \textit{minimal length} in the literature \cite{Kempf_Mangano_Mann,Brau,Fring_Gouba_Scholtz,Chang, Nozari_Etemadi,Maziashvili,Sprenger,Dey_Fring_Gouba, Dey_Fring_Acta,Dey_Fring_Khantoul,Maziashvili_Megrelidze,Dey_Thesis, Sobhani_Hassanabadi,Bhat_Dey_Faizal_Hou_Zhao,Lewis_Roman_Takeuchi, Falaye,Nascimento_Aguiar_Guedes}. An intimate connection between the gravitation and the existence of the fundamental length scale was proposed in \cite{Mead}. The minimal length has found to exist in string theory \cite{Veneziano}, loop quantum gravity \cite{Rovelli}, path integral quantum gravity \cite{Padmanabhan}, special relativity \cite{Amelino}, doubly special relativity \cite{Magueijo}, coherent states \cite{Trifonov_Review,Quesne_Penson_Tkachuk,Ghosh_Roy, Dey_Fring_squeezed,Dey_Fring_Gouba_Castro,Ching_Ng,Pedram,Dey,Dey_Hussin, Dey_Hussin_2,Dey_Fring_Hussin,Dey2,Fakhri_Hashemi,Ramirez_Reboiro,Jarvis_Lohe},, etc. Furthermore, some thought experiments \cite{Mead} in the spirit of black hole physics suggest that any theory of quantum gravity must be equipped with a minimum length scale \cite{Maggiore}, due to the fact that the energy required to probe any region of space below the Plank length is greater than the energy required to create a mini black hole in that region of space. In short, the existence of minimal measurable length, by now, has become a universal feature in almost all approaches of quantum gravity. For further informations on the subject one may follow some review articles devoted to the subject, for instance, \cite{Garay,Hossenfelder_Review}.

While NC theories have been proposed in a way that may circumvent the divergence problem in quantum gravity, the main obstacle is in the understanding of such theories experimentally. The effects of quantum gravity are expected to become relevant near the Planck length ($l_P\approx 10^{-35}m$) or at the energy scale near the Planck energy ($E_P\approx 10^{19} GeV$), which is about 15 orders of magnitude away from the energy range accessible to us today through the high energy scattering experiments. Astronomical observations have also failed to provide any promising evidence of quantum gravitational effects. It is, thus, beyond our ability to provide any experimental setup that could test quantum gravity. However, it has been shown recently that it is possible to test such theories by using an opto-mechancial experimental set-up \cite{Pikovski,DeyNPB}. Moreoever, if minimal length exists and its influence is severe in many directions of quantum gravity, the corresponding quantum mechanical structure has to be reformulated too. This gives rise the necessity to study the subject of NC quantum mechanics \cite{Gamboa,Girotti,Scholtz}. The best way to realize the effects of these deformations from its root is to study quantum optical models, which is what is our main motivation.

Let us now introduce a version of NC structure from a slightly different background and discuss about why this version is interesting for our purpose. We commence with a simple $q$-deformed oscillator algebra of the form
\begin{equation}\label{qDeformed}
A_qA_q^\dagger -q^2A_q^\dagger A_q=1, \qquad \vert q\vert < 1.
\end{equation}
The Fock space of the corresponding algebra \myref{qDeformed} can be defined by choosing $q$-deformed integers $[n]_q$ in such a way that the following relations hold
\begin{eqnarray}\label{qFock}
\vert n\rangle_q &:=& \frac{A_q^{\dagger n}}{\sqrt{[n]_q!}}\vert 0\rangle_q, \quad [n]_q!:= \displaystyle\prod_{k=1}^{n}[k]_q, \quad [0]_q! :=1, \notag\\
~[n]_q &:=& \frac{1-q^{2n}}{1-q^2}, \quad A_q\vert 0\rangle_q = 0, \quad ~_{q}\langle 0 \vert 0\rangle_q = 1.
\end{eqnarray}
It immediately follows that the operators $A_q$ and $A_q^\dagger$ act as lowering and raising operators, respectively, in the deformed Fock space
\begin{eqnarray}\label{qLadder}
A_q\vert n \rangle_q &=& \sqrt{[n]_q}~\vert n-1\rangle_q, \\
A_q^\dagger\vert n \rangle_q &=& \sqrt{[n+1]_q}~\vert n+1\rangle_q. \notag
\end{eqnarray}
It means that the states $\vert n\rangle_q$ form an orthonormal basis in the $q$-deformed Hilbert space $\mathcal{H}_q$ spanned by the vectors $\vert\psi\rangle :=\sum_{n=0}^{\infty}c_n\vert n\rangle_q$ with $c_n\in\mathbb{C}$, such that $\langle\psi\vert\psi\rangle = \sum_{n=0}^{\infty}\vert c_n\vert^2<\infty$. Therefore, the commutation relation between $A_q$ and $A_q^\dagger$ is realized as follows
\begin{equation}
[A_q,A_q^\dagger]=1+(q^2-1)A_q^\dagger A_q=1+(q^2-1)\hat{[n]}_q,
\end{equation}
where $\hat{[n]}_q=A_q^\dagger A_q$ is the number operator for the deformed system. The concept that the $q$-deformed algebras of type \myref{qDeformed} can be implemented for the construction of $q$-deformed harmonic oscillators was given by many authors \cite{Arik_Coon,Biedenharn}. Here we recall Refs. \cite{Bagchi_Fring,Dey_Fring_Gouba,Kempf_Mangano_Mann,Dey_Fring_Gouba_Castro} to construct a NCHO from the given algebra \myref{qDeformed}, instead. For this, we first express the deformed observables $X$ and $P$ in terms of the ladder operators $A_q, A_q^\dagger$ in the following form
\begin{equation}\label{XP}
X=\gamma(A_q^\dagger+A_q) \qquad \text{and} \qquad P=i\delta(A_q^\dagger-A_q).
\end{equation}  
Thereafter, by using \myref{qDeformed} we obtain the following commutation relation between the position and momentum variables \cite{Bagchi_Fring,Dey_Thesis}
\begin{equation}\label{NCOM}
[X,P]=\frac{4i\gamma\delta}{1+q^2}\left\{1+\frac{q^2-1}{4}\left(\frac{X^2}{\gamma^2}+\frac{P^2}{\delta^2}\right)\right\}.
\end{equation}
The interesting feature of such type of NC space-time \myref{NCOM} is that it leads to the existence of a minimal length as well as a minimal momentum \cite{Bagchi_Fring,Dey_Fring_Gouba}, which are also direct consequences of string theory. Furthermore, there exists a concrete self-adjoint representation of the ladder operators \cite{Dey_Fring_Gouba_Castro}
\begin{eqnarray}\label{HermitianRep}
A_q &=& \frac{i}{\sqrt{1-q^2}}\left(e^{-i\hat{x}}-e^{-i\hat{x}/2}e^{2\tau \hat{p}}\right), \\
A_q^\dagger &=& \frac{-i}{\sqrt{1-q^2}}\left(e^{i \hat{x}}-e^{2\tau \hat{p}}e^{i\hat{x}/2}\right), \notag
\end{eqnarray}
in terms of the canonical coordinates $x,p$ satisfying $[x,p]=i\hbar$, with $\hat{x}=x\sqrt{m\omega/\hbar}$ and $\hat{p}=p/\sqrt{\hbar m\omega}$  being dimensionless observables, and the deformation parameter $q$ being parametrized to $q=e^\tau$. It follows that the observables \myref{XP}, which satisfy \myref{NCOM} are Hermitian with respect to the representation \myref{HermitianRep}, i.e. $X^\dagger =X, P^\dagger = P$. As obviously, our representation \myref{HermitianRep} is not unique and with further investigations it may be possible to find other Hermitian representations. However, for our purpose it is important that there exists at least one such representation providing a self-consistent description of a physical system.
Nevertheless, by imposing the constraint on the parameters $\gamma\delta=\hbar/2$ in \myref{NCOM} and, assuming the deformation parameter $q$ to be of the form $q=e^{2\check{\tau}\beta^2}$ followed by a nontrivial limit $\beta\rightarrow 0$, we obtain \cite{Bagchi_Fring,Dey_Thesis}
\begin{equation}\label{NCOMCommutator}
[X,P]=i\hbar(1+\check{\tau}P^2),
\end{equation} 
where $\check{\tau}=\tau/(m\omega\hbar)$ has the dimension of inversed squared momentum, with $\tau\in\mathbb{R}+$ being dimensionless. Note that in the limit $\tau\rightarrow 0$, i.e. for $q\rightarrow 1$, the commutation
relation \myref{NCOMCommutator} reduces to the usual canonical commutation relation. For further informations on this version of NC structure, one may follow \cite{Bagchi_Fring,Dey_Fring_Gouba,Dey_Fring_Khantoul,Dey_Thesis}, where one can find elaborate discussions on the physical implications of the algebra. A simple representation of the commutator \myref{NCOMCommutator} in terms of the canonical observables $x,p$ satisfying $[x,p]=i\hbar$ is given by
\begin{equation}\label{NHREP}
X=(1+\check{\tau}p^2)x, \quad P=p,
\end{equation}
which is, however, non-Hermitian with respect to the standard inner product. There are several other non-Hermitian as well as some Hermitian representations of the algebra \myref{NCOMCommutator} \cite{Dey_Fring_Khantoul}, nevertheless, we will mainly consider the non-Hermitian representation \myref{NHREP} and discuss how to deal with these kind of difficulties for our purpose. Note that, in order to build quantum optical models, like coherent and nonclassical states, we may either consider the  $q$-deformed version satisfying \myref{qLadder} or, we can take the observables $X$ and $P$ represented in \myref{NHREP} directly to build a meaningful Hamiltonian, for example, a NC harmonic oscillator (NCHO)
\begin{equation}\label{NCHOHam}
H=\frac{P^2}{2m}+\frac{m\omega^2}{2}X^2-\hbar\omega \left(\frac{1}{2}+\frac{\tau}{4}\right).
\end{equation}
Here the ground state energy is conventionally shifted to allow for a factorization of the energy. Nevertheless, both of the approaches are equivalent as well as interrelated to each other and represent a non-Hermitian and NC system.
\subsection{PT-symmetry}\label{SubSec32}
Hermiticity is a property of quantum operators that ensures real eigenvalues as well as unitary time evolution when this operator is taken to be the Hamiltonian. Therefore, in order to construct a meaningful physical system, Hermiticity is a property that is desirable. However, by now, it is also well-established that Hermiticity is not necessary and the non-Hermitian Hamiltonians could play an important role in the formulation of complete and fundamental quantum theories \cite{Bender_Boettcher,Bender_Making_Sense}. The possibility that those systems can possess discrete eigenstates with real positive energies was indicated by von Neumann and Wigner \cite{neumann} almost eighty five years ago. Later, this type of systems were under more intense scrutiny and, nowadays, the properties of these so-called BICs (bound states in the continuum) are fairly well-understood for many concrete examples \cite{friedrich1} together with their bi-orthonormal eigenstates \cite{persson}.

Whereas, the above type of Hamiltonians only possess single states with these ``strange properties" \cite{neumann}, it was observed fairly recently in a ground-breaking numerical study by Bender and Boettcher \cite{Bender_Boettcher} that the Hamiltonians with potential terms $V = x^2\left(ix\right)^\nu$ for $\nu \geq 0$ possess entirely real and positive spectra. It was argued that the reality of spectrum is guaranteed if the Hamiltonian is symmetric under the simultaneous operation of parity $\mathcal{P}$ and time-reversal $\mathcal{T}$ operators, such that $\mathcal{PT}:~x\rightarrow -x, p\rightarrow p, i\rightarrow -i$. However, later it became evident that all what is required for the reality of spectrum is an anti-linear symmetry \cite{wigner1} and, $\mathcal{PT}$ is one of the examples only. In fact, when the wavefunctions are simultaneous eigenstates of the Hamiltonian and the $\mathcal{PT}$-operator, one can easily argue that the spectrum has to be real \cite{wigner1,Bender_Making_Sense,bender_brody_jones}. However despite the fact that $\left[\mathcal{PT},H\right]=0$, this is not always guaranteed, because the $\mathcal{PT}$ operator is an anti-linear operator \cite{weigert}. As a consequence one may also encounter conjugate pair of eigenvalues for broken $\mathcal{PT}$ symmetry \cite{bender_brody_jones}, that is when $[\mathcal{PT},H]=0$ but $\mathcal{PT}\phi\neq \phi$. One may use various techniques \cite{dorey} to verify case-by-case, whether the $\mathcal{PT}$ symmetry is broken or not.

Therefore a $\mathcal{PT}$-symmetric Hamiltonian, though it is non-Hermitian, in principle in the $\mathcal{PT}$ unbroken regime can also produce a quantum theory similar to a Hermitian Hamiltonian. However, what is essential is to have a fully consistent quantum theory whose dynamics is described by a non-Hermitian Hamiltonian. In order to achieve this, one needs to modify the inner product for the corresponding Hilbert space. The natural choice of the inner product suitable for $\mathcal{PT}$-symmetric quantum mechanics is the $\mathcal{PT}$-inner product which can be defined as
\begin{equation}
\langle \phi \vert \psi \rangle ^{\mathcal{PT}}=\int \left[\phi\left(x\right)\right]^{\mathcal{PT}}\psi\left(x\right) dx=\int \left[\phi\left(-x\right)\right]^\ast \psi \left(x\right)dx.
\end{equation}
Note that, the boundary conditions (vanishing $\phi$, $\psi$ at $x\rightarrow\pm\infty$) must be imposed properly at this point to solve the eigenfunctions of the time independent Schr{\"o}dinger equation, which are located in this context within the wedges bounded by Stokes lines in the complex $x$-plane \cite{Bender_Boettcher} and that is the reason why one must integrate the system within this specified region. However, the inner product is not yet acceptable to formulate a valid quantum theory, because the norm of a state is not always positive definite, which is once again due to the fact that the wavefunctions may not be simultaneous eigenfunctions of $H$ and $\mathcal{PT}$ due to the antilinearity property of the $\mathcal{PT}$ operator. Bender, Brody and Jones \cite{bender_brody_jones} overcame this problem consistently by introducing a $\mathcal{CPT}$-inner product, which was later studied by many people; see, for instance \cite{weigert_completeness,bender_brody_jones_prd,Bebiano}.
\subsection{Pseudo-Hermiticity}\label{SubSec33}
The concept of pseudo-Hermiticity was introduced very early in 1940s by Dirac and Pauli \cite{pauli}, and was discussed later by Lee, Wick, and Sudarshan \cite{sudarshan,Lee_Wick}, who were trying to resolve the problems that arose in the context of quantizing electrodynamics and other quantum field theories in which negative norm states appear as a consequence of renormalization. Even before the discovery of $\mathcal{PT}$-symmetry \cite{Bender_Boettcher} and the introduction of the $\mathcal{CPT}$-inner product, there have been very general considerations \cite{dieudonne,Scholtz_Geyer_Hahne} addressing the question of how a consistent quantum mechanical framework can be constructed from the non-Hermitian Hamiltonian systems. It was understood at that time that quasi-Hermitian systems would lead to positive inner products. The concept was illustrated later by Mostafazadeh \cite{mostafazadeh1}, who proposed that instead of considering quasi-Hermitian Hamiltonians one may investigate pseudo-Hermitian Hamiltonians satisfying
\begin{eqnarray}\label{hermiticity}
&& h=\eta H \eta^{-1}=h^\dagger=\eta^{-1}H^\dagger \eta, \\
&& H^\dagger=\rho H \rho^{-1}~~\text{with}~~\rho=\eta^\dagger\eta, \notag
\end{eqnarray}
where $\rho$ is a linear, invertible, Hermitian and positive operator acting in the Hilbert space, such that $H$ becomes a self-adjoint operator with regard to this metric $\rho$, as explained in more detail below. $\eta$ is often called the Dyson map \cite{dyson}. Note that the usual Hermiticity condition is recovered with the choice of $\eta$ to be $\bf{1}$. Since the Hermitian Hamiltonian $h$ and non-Hermitian Hamiltonian $H$ are related by a similarity transformation, they belong to the same similarity class and, therefore, have the same eigenvalues. The time-independent Schr{\"o}dinger equations corresponding to the Hermitian and non-Hermitian Hamiltonian are
\begin{equation}
h\phi=\epsilon \phi \qquad \text{and} \qquad H\Phi=\epsilon \Phi,
\end{equation}
respectively, where the wavefunctions are related as
\begin{equation}
\Phi=\eta^{-1} \phi.
\end{equation}
Therefore, the inner products for the wavefunctions $\Phi$ related to the non-Hermitian $H$ may now simply taken to be
\begin{equation}\label{innerproduct}
\langle\Phi\vert\Phi'\rangle_\eta:=\langle\Phi\vert\eta^2\Phi'\rangle,
\end{equation}
where the inner product on the right hand side of \myref{innerproduct} is the conventional inner product associated to the Hermitian Hamiltonian $h$. Crucially we have $\langle\Phi\vert H \Phi'\rangle_\eta=\langle H \Phi\vert\Phi'\rangle_\eta$.

To summarize, it is conceptually straight forward to compute the Hermitian counterpart $h$ of the non-Hermitian Hamiltonian $H$, for which one needs to construct the metric operator followed by the equation \myref{hermiticity}. Thus a key task that remains to calculate in this approach is to find $\rho$ and $\eta$. In practical terms, however, there are very few examples \cite{bagchi_quesne_roychoudhury,znojil_geyer} where one can compute them in an exact manner, as for example; see, \cite{Dey_Fring_Mathanaranjan,Dey_Fring_Mathanaranjan1} for an exact form of the metric which was derived in the context of Euclidean Lie algebraic Hamiltonians. However, there are many other methods such as spectral theory, perturbation technique  \cite{bender_brody_jones_prd,ghatak_mandal}, Moyal product approach \cite{faria_fring1} etc., which one may follow for the construction of the metric operator.

Meanwhile there are numerous experimental conformations of effects and
applications resulting from the presence of $\mathcal{PT}$-symmetry in many
branches of physics. Most notably are the applications in optics exploiting
the formal analogy between the stationary Schr\"odinger equation and the
Helmholtz equation describing monochromatic linearly polarized light. This
has led to the production of material with controllable gain and loss \cite{Guo,ruter2010,regensburger}, which led for instance to the discovery of
meta-materials \cite{feng2013} with unidirectional invisibility \cite{lin2011} or complete absorption ability for incoming radiation referred to as coherent perfect absorbers \cite{chong2010}. Further applications of $\mathcal{PT}$-symmetric systems can be found in the
stimulation of superconductivity \cite{chtchelkatchev}, microwave cavities
physics \cite{bittner2012} and in nuclear magnetic resonance quantum systems 
\cite{zheng2013}. For more informations in this regard one may refer to some books on the subject \cite{Moiseyev_Book,Bagarello_Book}.
\section{\textbf{Coherent states for non-Hermitian systems}}\label{sec4}
Let us now construct the coherent states for the non-Hermitian systems described in the previous section. Here, we discuss several different types of coherent states arising from this scenario. We will mainly focus here on examples for the NCHO \myref{NCHOHam}, however, by following our procedure it is possible to construct coherent states for any other models. As discussed before in Sec. \ref{SubSec31}, the Hamiltonian \myref{NCHOHam} is non-Hermitian, since, the position and momentum operators $X,P$ are not self adjoint and satisfy \myref{NHREP}. But, we can use the standard techniques of non-Hermtitian systems to find real eigenvalues of the system as described in Secs. \ref{SubSec32} and \ref{SubSec33}. In particular, we compute the isospectral Hermitian counterpart $h$ of the non-Hermitian Hamiltonian $H$ \myref{NCHOHam} by utilizing the pseudo Hermiticity property \myref{hermiticity} as follows \cite{Dey_Fring_squeezed}
\begin{eqnarray}\label{HermHam}
h &=& \eta H \eta^{-1} \\
&=& \frac{p^2}{2m}+\frac{m\omega^2}{2}x^2+\frac{\omega\tau}{4\hbar}\left(p^2x^2+x^2p^2+2xp^2x\right) \notag\\
&& -\hbar\omega\left(\frac{1}{2}+\frac{\tau}{4}\right)+\mathcal{O}(\tau^2)~, \notag
\end{eqnarray}
with the metric $\eta=(1+\check{\tau}p^2)^{-1/2}$. We consider a perturbative treatment here and decompose the above Hamiltonian \myref{HermHam} as $h=h_0+h_1$. Now, taking $h_0$ to be the standard harmonic oscillator and following the common techniques of Rayleigh-Schr\"odinger perturbation theory, the energy eigenvalues of $H$ and $h$ are computed \cite{Kempf_Mangano_Mann,Dey_Fring_Gouba,Dey_Fring_squeezed} to lowest order to
\begin{equation}\label{fn}
E_n=\hbar\omega nf^2(n)=\hbar\omega (A n+B n^2)+\mathcal{O}(\tau^2)~,
\end{equation}
with $A=(1+\tau/2)$ and $B=\tau/2$. The corresponding eigenstates are
\begin{eqnarray}\label{eigenstates}
\vert\phi_n\rangle &=& \vert n\rangle-\frac{\tau}{16}\sqrt{(n-3)^{(4)}}\vert n-4\rangle \\
&& +\frac{\tau}{16}\sqrt{(n+1)^{(4)}}\vert n+4\rangle +\mathcal{O}(\tau^2)~, \notag
\end{eqnarray} 
where $Q^{(n)}:=\prod_{k=0}^{n-1}(Q+k)$ denotes the Pochhammer symbol with the raising factorial. In what follows, we will drop the explicit mentioning of the order in $\tau$, understanding that all our computations are carried out to first order. Having obtained all the prerequisites, we can now construct the coherent states for the NCHO \myref{NCHOHam}. 
\subsection{Nonlinear coherent states}\label{SubSec41}
In order to construct the nonlinear coherent states (NLCS), let us start by considering a set of generalized ladder operators $A^\dagger$ and $A$ in terms of the bosonic creation and annihilation operators $a^\dagger$ and $a$
\begin{eqnarray}\label{genladder}
A^\dagger &=& f(\hat{n})a^\dagger=a^\dagger f(\hat{n}+1),\\
A &=& af(\hat{n})=f(\hat{n}+1)a,\notag
\end{eqnarray}
where $f(n)$ is an operator-valued function of the Hermitian number operator $\hat{n}=a^\dagger a$. The operators $A$ and $A^\dagger$ therefore obey the following nonlinear commutator algebras
\begin{eqnarray}\label{defcommutator}
&& \big[A,A^\dagger\big]=(\hat{n}+1)f^2(\hat{n}+1)-\hat{n}f^2(\hat{n}), \\
&& \big[\hat{n},A\big]=-A, \quad \big[\hat{n},A^\dagger\big]=A^\dagger, \notag
\end{eqnarray} 
where the nonlinearity arises from $f(\hat{n})$. Clearly, with the choice of $f(\hat{n})=1$, the deformed algebra \myref{defcommutator} reduces to the Heisenberg algebra
\begin{equation}
\big[a,a^\dagger\big]=1, \quad \big[\hat{n},a\big]=-a \quad \text{and} \quad \big[\hat{n},a^\dagger\big]=a^\dagger.
\end{equation}
In analogy to the Glauber states \cite{Glauber}, the NLCS \cite{Filho_Vogel,Manko_Marmo_Sudarshan_Zaccaria,Sivakumar} are, therefore, defined as the right eigenvector of the generalized annihilation operator $A$:
\begin{equation}\label{defeigen}
A\big\vert\alpha,f\big\rangle=\alpha\big\vert\alpha,f\big\rangle,
\end{equation}
where $\alpha$ is a complex eigenvalue, which is however to be expected as $A$ is non-Hermitian. Solving the eigenvalue equation \myref{defeigen} one then obtains an explicit expression of coherent state in number state representation \cite{Dey_Fring_Hussin}
\begin{eqnarray}\label{noncoherent}
&& \big\vert\alpha,f\big\rangle=\frac{1}{\mathcal{N}(\alpha,f)}\displaystyle\sum_{n=0}^{\infty}\frac{\alpha^n}{\sqrt{n!}~h(n)}\vert n\rangle, \qquad \alpha\in\mathbb{C}, \\
&& \text{where}~~~h(n)=\left\{ \begin{array}{lcl}
1 & \mbox{if}
& n=0 \\ \displaystyle\prod_{k=0}^{n} f(k) & \mbox{if} & n>0~.\end{array}\right. \notag
\end{eqnarray}
It is possible to define another set of ladder operators $B$ and $B^\dagger$  \cite{Roy_Roy}, which are canonically conjugate to $A$ and $A^\dagger$
\begin{equation}
B^\dagger=a^\dagger\frac{1}{f(\hat{n})} \qquad \text{and} \qquad B=\frac{1}{f(\hat{n})}a,
\end{equation}
so that one can easily check $[A,B^\dagger]=[B,A^\dagger]=1$, which allows one to write the displacement operator
\begin{equation}
D\big(\alpha,f\big)=e^{\alpha B^\dagger-\alpha^\ast A},
\end{equation}
and construct NLCS through
\begin{equation}
\big\vert\alpha,f\big\rangle=D\big(\alpha,f\big)\vert 0\rangle~.
\end{equation}
The outcome coincides exactly with \myref{noncoherent}. The normalization constant can be computed from the requirement $\big\langle \alpha,f\big\vert\alpha,f\big\rangle=1$, so that
\begin{equation}\label{normalisation}
\mathcal{N}^2(\alpha,f)=\displaystyle\sum_{n=0}^{\infty}\frac{\vert\alpha\vert^{2n}}{n!~ h^2(n)}~~.
\end{equation}
Since, the eigenfunctions in our case are given by \myref{eigenstates}, the NLCS for our case turns out to be
\begin{equation}\label{NcCoherent}
\vert\alpha,f,\phi\rangle = \frac{1}{\mathcal{N}(\alpha,f)}\displaystyle\sum_{n=0}^{\infty}\frac{\alpha^n}{\sqrt{n!}f(n)!}\vert \phi_n\rangle, \qquad  \alpha\in \mathbb{C}~,
\end{equation}
which when rewritten in terms of the Fock states become
\begin{equation}\label{NcCoherentAlt}
\vert\alpha,f,\phi\rangle = \frac{1}{\mathcal{N}(\alpha,f)}\displaystyle\sum_{n=0}^{\infty}\frac{\mathcal{C}(\alpha,n)}{\sqrt{n!}f(n)!}\vert n\rangle,
\end{equation}
where
\begin{eqnarray}
f^2(n)!&=&\frac{\tau^n}{2^n}\left(2+\frac{2}{\tau}\right)^{(n)}, \\
\frac{1}{f^2(n)!} &=& 1-\frac{\tau}{4}n(3+n)+\mathcal{O}(\tau^2), \\
\mathcal{N}^2(\alpha,f) &=& \displaystyle\sum_{n=0}^{\infty}\frac{\vert\alpha\vert^{2n}}{n!f^2(n)!} \label{NAF}\\
&=& e^{\vert\alpha\vert^2}\left(1-\tau\vert\alpha\vert^2-\frac{\tau}{4}\vert\alpha\vert^4\right)+\mathcal{O}(\tau^2), \notag
\end{eqnarray}
 and
\begin{eqnarray}\label{DisplacedCoh}
&&\mathcal{C}(\alpha,n)= \\
&&\left\{ \begin{array}{ll}
\alpha^n-\frac{\tau}{16}\alpha^{n+4}\frac{f(n)!}{f(n+4)!},~~~~0\leq n \leq 3
& \\ \alpha^n-\frac{\tau}{16}\alpha^{n+4}\frac{f(n)!}{f(n+4)!}+\frac{\tau}{16}\alpha^{n-4}\frac{n!}{(n-4)!}\frac{f(n)!}{f(n-4)!}, & n\geq 4~.\end{array}\right. \notag
\end{eqnarray}
We will use these states to compute various physical properties, but let us first elaborate further on their mathematical consistency.
\subsubsection{Resolution of identity}
The resolution of identity is an important mathematical property that a coherent state must possess. The vectors $\vert\alpha,f\rangle$ in \myref{noncoherent} are mathematically well-defined in the domain $\mathcal{D}$ of allowed $\vert\alpha\vert^2$ for which the series \myref{normalisation} converges. The range of $\vert\alpha\vert^2,~0\leq\vert\alpha\vert^2<R$, is determined by the radius of convergence $R=\lim_{n\rightarrow\infty}\sqrt{\rho_n}$, which may be finite or infinite depending on the behavior of $\rho_n$ for large $n$. Here $\rho_n=\prod_{k=1}^{n} e_k=n!f^2(n)!$, $\rho_0=1$ and $e_n=nf^2(n)$ being an infinite sequence of positive numbers, with $e_0=0$. Therefore, a family of such coherent states \myref{noncoherent} is an \textit{overcomplete} set of vectors in a Hilbert space $\mathcal{H}$, labeled by a continuous parameter $\alpha$ which belongs to a complex domain $\mathcal{D}\subset\mathbb{C}$ (For $R=\infty$, $\mathcal{D}=\mathbb{C}$). To be more precise, since $\vert n\rangle$ forms an orthonormal basis in the Hilbert space $\mathcal{H}$, the vectors $\vert\alpha,f\rangle$ must satisfy the resolution of identity (completeness relation) with a weight function $\Omega$
\begin{eqnarray}\label{completeness}
\int\int_{\mathcal{D}} \frac{\mathcal{N}(\alpha,f)}{\pi}\vert\alpha,f\rangle\langle\alpha,f\vert~\Omega(\vert\alpha\vert^2)~d^2 \alpha = \mathbb{I}_\mathcal{H}.
\end{eqnarray}
By considering $\alpha=re^{i\theta}$, the left hand side of \myref{completeness} turns out to be
\begin{eqnarray}
&& \displaystyle\sum_{m,n=0}^\infty\frac{1}{2\pi\sqrt{\rho_m\rho_n}}\int_{0}^Rr^{m+n}\Omega(r^2)d(r^2) \\
&& \times \int_{0}^{2\pi}e^{i\theta(m-n)}d\theta~\vert m\rangle\langle n\vert =\displaystyle\sum_{n=0}^\infty\frac{1}{\rho_n}\int_{0}^{R}t^n\Omega(t)dt~\vert n\rangle\langle n\vert, \notag
\end{eqnarray}
such that one ends up with an infinite set of constraints
\begin{eqnarray}\label{measure}
\int_0^R t^{n}\Omega(t)dt = \rho_n, \qquad 0<R\leq \infty,
\end{eqnarray}
for which the completeness relation \myref{completeness} holds. Therefore, one can construct the coherent states \myref{noncoherent} for any model corresponding to a known $f(n)$, provided that there exists a measure $\Omega(t)$ which satisfies \myref{measure}. The explicit expression of the measure can be found, first, by associating \myref{measure} with the classical moment problem, where $\rho(n)>0$ are the power moments of the unknown function $\Omega(t)>0$ and, subsequently, by carrying out the integration by using the standard techniques of the Mellin transforms \cite{Oberhettinger}. In our case 
\begin{equation}\label{NCHOCS}
\rho_n=n!f^2(n)!=\left(\frac{\tau}{2}\right)^n\frac{n!(n+\frac{2}{\tau}+1)!}{(1+\frac{2}{\tau})!},
\end{equation}
such that we obtain the accurate expression of the Borel measure $\Omega(t)$ as follows \cite{Dey2}
\begin{equation}\label{measure1}
\Omega(t)=\frac{2^{\frac{1}{2}(4+\mu+\beta)}}{\tau\Gamma(1+\beta)}\left(\frac{t}{\tau}\right)^{\frac{\mu+\beta}{2}}K_{\mu-\beta}(2\sqrt{\frac{2t}{\tau}}).
\end{equation}
This establishes that a measure leading to \myref{NCHOCS} exists and can indeed be found explicitly.
\subsubsection{Nonclassical properties}\label{SebSec42}
In order to study the behavior of the NLCS \myref{noncoherent}, we first evaluate the expectation values of the nonlinear quadrature operators $Y,Z$ and their squares $Y^2,Z^2$, so that we obtain \cite{Dey_Fring_Hussin}
\begin{equation}
\Delta Y^2 = R+\tau\left(\frac{1}{4}+\frac{\vert\alpha\vert^2}{2}\right), \quad \Delta Z^2 = R-\tau\left(\frac{1}{4}+\frac{\vert\alpha\vert^2}{2}\right).
\end{equation}
The right hand side of the generalized uncertainty relation (GUR)
\begin{equation}\label{GUR}
\Delta Y\Delta Z\geq\frac{1}{2}\Big\vert\langle\alpha,f,\Phi\vert [Y,Z]\vert\alpha,f,\Phi\rangle_\eta\Big\vert,
\end{equation}
\begin{figure*}
\centering   \includegraphics[width=9.0cm,height=6.2cm]{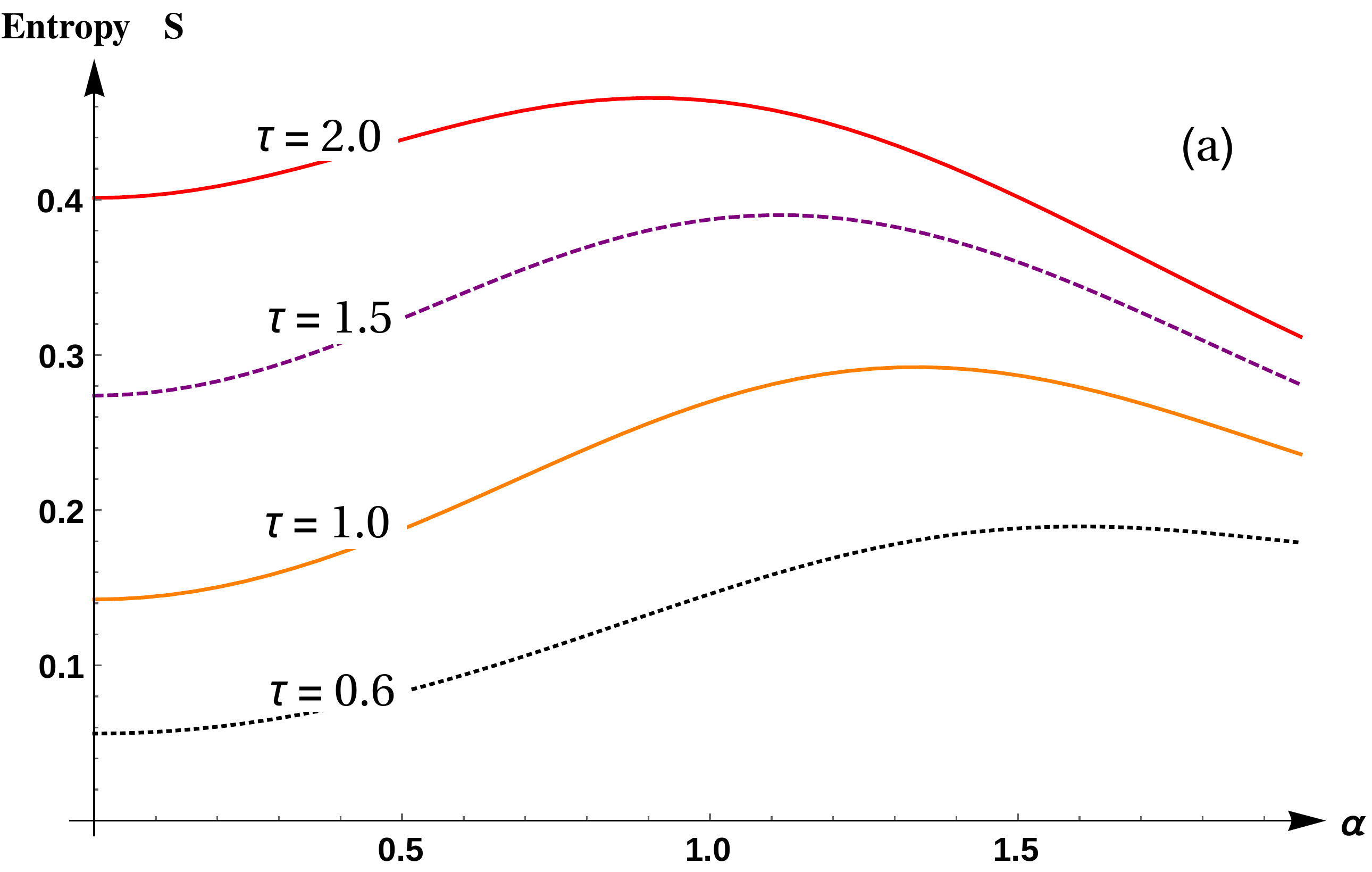}
\includegraphics[width=9.0cm,height=6.2cm]{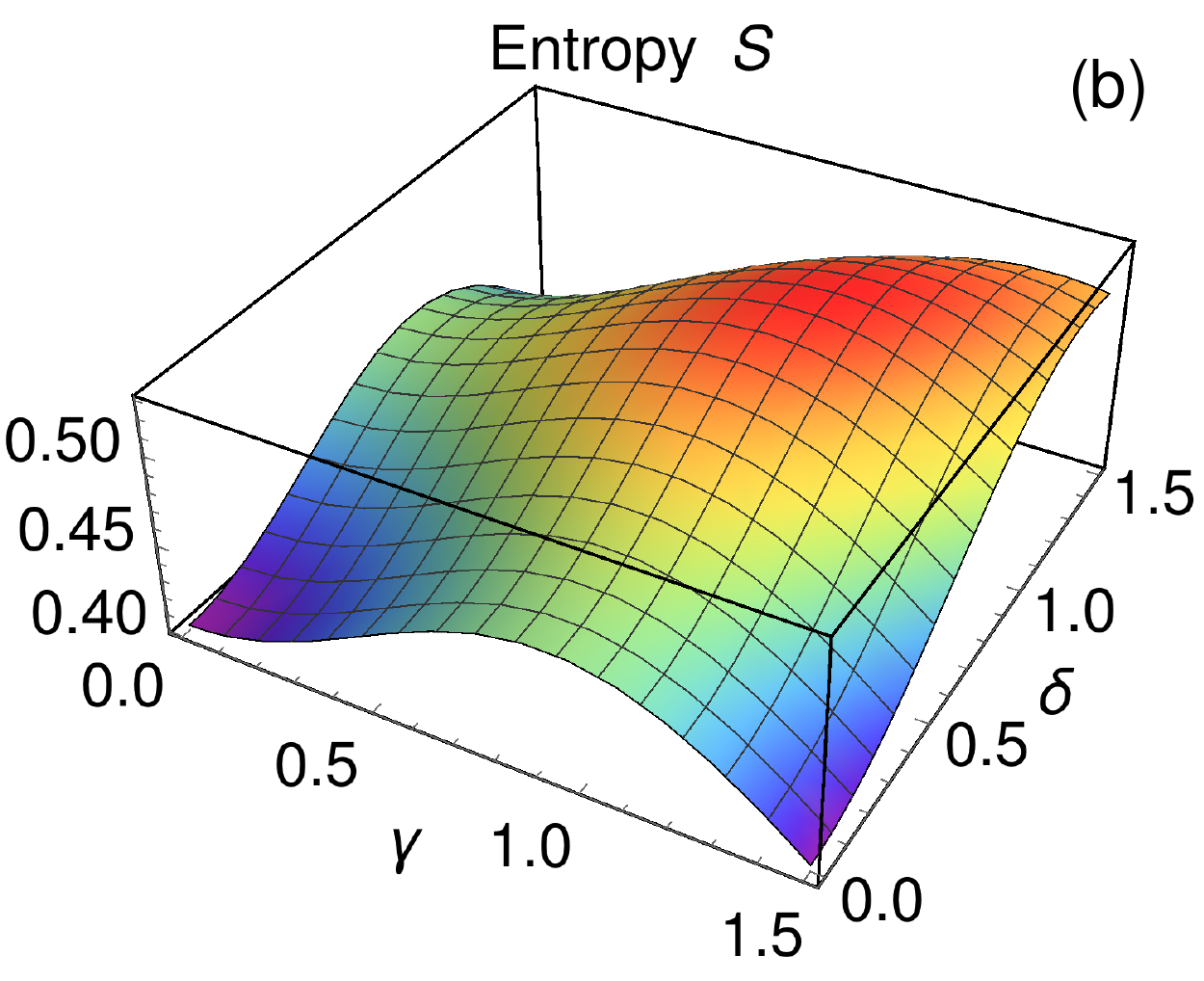}
\caption{\small{Linear entropy of NC NLCS (a) for different values of $\tau$, (b) in the complex plane of $\alpha$, with $\alpha =\gamma+i\delta$ and $\tau =2$ \cite{Dey_Hussin}.}}
\label{fig1}
\end{figure*}
is computed to
\begin{equation}\label{RHSGUR}
R=\frac{1}{2}\Big[1+\tau\langle\alpha,f,\Phi\vert Z^2\vert\alpha,f,\Phi\rangle_\eta\Big]=\frac{1}{4}\Big[2+\tau-\tau(\alpha-\alpha^\ast)^2\Big].
\end{equation}
It is, thus, easy to check that the GUR \myref{GUR} is saturated in this case; i.e. $\Delta Y\Delta Z =R$ and, therefore, the coherent states in NC space can be referred as intelligent states. However, unlike the coherent states of the ordinary harmonic oscillator, uncertainties in two quadratures in this case are not equal to each other. Rather, the quadrature $Z$ is squeezed below the right hand side of the uncertainty relation $R$ \myref{RHSGUR}, whereas the quadrature $Y$ is expanded correspondingly, such that the uncertainty relation saturates. Therefore, the state belongs to the family of ideal squeezed states. On the other hand, while we study the photon number squeezing properties the Mandel parameter \myref{Mandel} turns out to be negative, $Q=-\tau\vert\alpha\vert^2/2$, suggesting a sub-Poissonian statistics and the state is number squeezed.
\paragraph{Beam splitter entanglement:}\label{SubBeam}
The nonclassical nature of NC coherent states is quite obvious from the above analysis. Here, we would like to verify our results by using the quantum beam splitter, which was described in Sec. \ref{SubSec25}. We consider the NC coherent states \myref{NcCoherentAlt} as one of the inputs, while a vacuum state $\vert 0\rangle$ at the other. The output states in this case are computed to 
\begin{eqnarray}\label{NcCoherentInput}
\vert\text{out}\rangle &=& \mathcal{B}\big(\vert\alpha,f,\phi\rangle\otimes\vert 0\rangle\big) \\
&=& \frac{1}{\mathcal{N}(\alpha,f)}\displaystyle\sum_{n=0}^{\infty}\frac{\mathcal{C}(\alpha,n)}{\sqrt{n!}f(n)!}\mathcal{B}\big(\vert n\rangle\otimes\vert 0\rangle\big). \notag
\end{eqnarray}
Substituting \myref{BeamFock} in \myref{NcCoherentInput} and following the similar steps as in \cite{Dey_Hussin}, we compute the reduced density matrix of the output states, such that the linear entropy becomes
\begin{eqnarray}\label{Entropy}
&& S = 1- \\
&& \frac{1}{\mathcal{N}^4(\alpha,f)} \displaystyle\sum_{q=0}^\infty\displaystyle\sum_{s=0}^\infty\displaystyle\sum_{m=0}^{\infty-\text{max}(q,s)} \displaystyle\sum_{n=0}^{\infty-\text{max}(q,s)} \vert t\vert^{2(q+s)}\vert r\vert^{2(m+n)} \notag \\ 
&& \times\frac{\mathcal{C}(\alpha,m+q)\mathcal{C}^\ast(\alpha,m+s)\mathcal{C}(\alpha,n+s)\mathcal{C}^\ast(\alpha,n+q)}{q!s!m!n!f(m+q)!f(m+s)!f(n+s)!f(n+q)!}~. \notag
\end{eqnarray}
Assuming the input states of the beam splitter to be nonclassical, we expect the output states to be entangled and, hence, a finite amount of linear entropy must be created. The results demonstrated in Fig. \ref{fig1} confirm our expectations and establishes the nonclassical nature of the NC coherent states. In contrast, when $f(n)=1$, which corresponds to the case of ordinary harmonic oscillator, the output states are not entangled and naturally we do obtain a null entropy. The most interesting fact is that when we enhance the noncommutativity by increasing the value of the parameter $\tau$, the entanglement rises accordingly and, therefore, becomes more and more nonclassical as shown in the left panel of Fig. \ref{fig1}.
\subsection{$q$-deformed coherent states} \label{SubSec42}
In order to construct the $q$-deformed coherent states we can simply use the eigenvalue definition \myref{eigen} with the usual annihilation operator $a$ being replaced by that of the $q$-deformed system $A_q$ \myref{qLadder}, so that we end up with \cite{Dey}
\begin{equation}\label{qdefcoherent}
\big\vert\alpha\big\rangle_q=\frac{1}{\mathcal{N}_q(\alpha)}\displaystyle\sum_{n=0}^{\infty}\frac{\alpha^n}{\sqrt{[n]_q!}}\vert n\rangle_q \quad \alpha\in\mathbb{C},
\end{equation}
with the normalization constant being represented in terms of the $q$-deformed exponential $E_q(\vert\alpha\vert^2)$ as follows
\begin{equation}\label{NormqD}
\mathcal{N}_q^2(\alpha)=\displaystyle\sum_{n=0}^{\infty}\frac{\vert\alpha\vert^{2n}}{[n]_q!}=E_q\big(\vert\alpha\vert^2\big).
\end{equation}
However, the interesting fact is that the coherent states emerging from the $q$-deformed structure \myref{qdefcoherent} coincides with that of the nonlinear system \myref{noncoherent} for
\begin{equation}
f(n)=\sqrt{[n]_q/n}.
\end{equation}
Let us now analyze the state and find out whether the state is classical-like or whether it possesses nonclassical properties. Straightforward computations of the expectation values of the deformed quadratures $Y_q$ and $Z_q$, yields   \cite{Dey}
\begin{eqnarray}
(\Delta Y_q)^2\Big\vert_{\vert\alpha\rangle_q}
&=& (\Delta Z_q)^2\Big\vert_{\vert\alpha,f\rangle_q} \notag \\ 
&=& \frac{1}{2}\Big\vert{}_{q}\!\big\langle \alpha\big\vert [Y_q,Z_q]\big\vert \alpha\big\rangle_q\Big\vert \notag \\
&=& \frac{1}{4}\Big\{1+\left(q^2-1\right)\vert\alpha\vert^2\Big\},
\end{eqnarray}
so that the GUR
\begin{eqnarray}\label{qGUR}
\Delta Y_q~\Delta Z_q\Big\vert_{\vert\alpha\rangle_q} &\geq & \frac{1}{2}\Big\vert{}_q\big\langle \alpha\big\vert [Y_q,Z_q]\big\vert \alpha\big\rangle_q\Big\vert,
\end{eqnarray}
is saturated in this case with uncertainties of the two quadratures $Y_q$ and $Z_q$ being identical to each other. The coherent states $|\alpha\rangle_q$ are, therefore, intelligent states and carry the information of a well-behaved coherent states \cite{Dey}, like the Glauber coherent states \myref{GlauberCoherent}. Next we compute the average of the number operator and its dispersion to obtain the Mandel parameter as follows \cite{Dey}
\begin{eqnarray}\label{qdefMandel}
Q_q=(q^2-1)\left\vert\alpha\right\vert^2 \qquad q\leq 1.
\end{eqnarray}
Note that in the case when $q=1$, which corresponds to the Glauber coherent states, the Mandel parameter becomes zero. Which means, for ordinary coherent states the photon distribution is always Poissonian and, therefore, the number squeezing is absent in that case \cite{Gerry_Knight_Book}. However, if one considers the $q$-deformed case \myref{qdefMandel} and restricts $-1<q<1$ further, the photon distribution remains sub-Poissonian.
\subsection{Gazeau-Klauder coherent states} \label{SubSec43}
The Gazeau-Klauder (GK) coherent states \cite{Gazeau_Klauder,Antoine_Gazeau_Monceau_Klauder_Penson} for a Hermitian Hamiltonian $h$ with discrete bounded below and nondegenerate eigenspectrum are defined as a two parameter set 
\begin{equation}
\left\vert J,\gamma ,\phi \right\rangle =\frac{1}{\mathcal{N}(J)}%
\sum\limits_{n=0}^{\infty }\frac{J^{n/2}\exp (-i\gamma e_{n})}{\sqrt{\rho
_{n}}}\left\vert \phi _{n}\right\rangle ,  \label{GK}
\end{equation}
\begin{figure*}
\centering   \includegraphics[width=9.0cm,height=6.2cm]{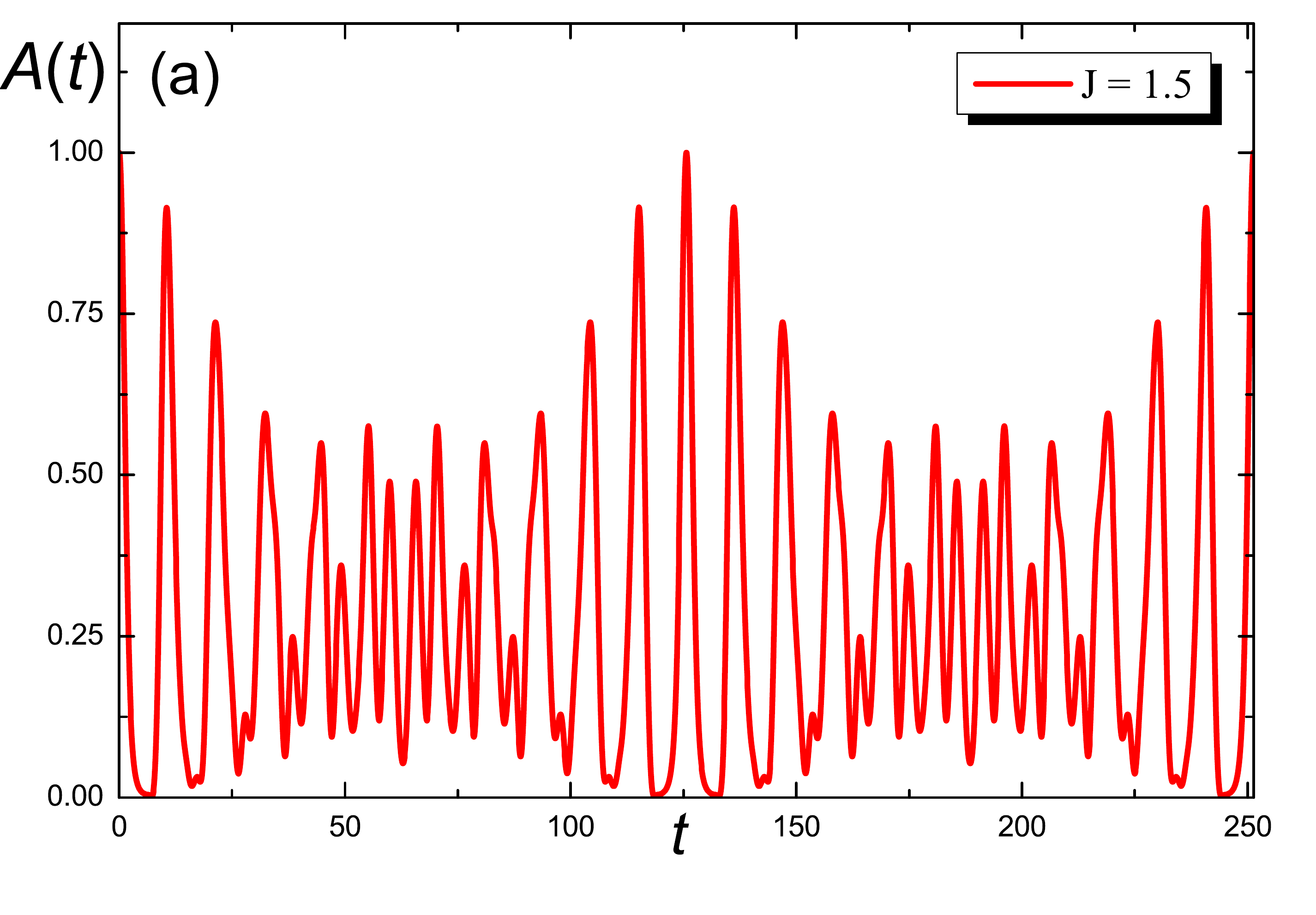}
\includegraphics[width=9.0cm,height=6.2cm]{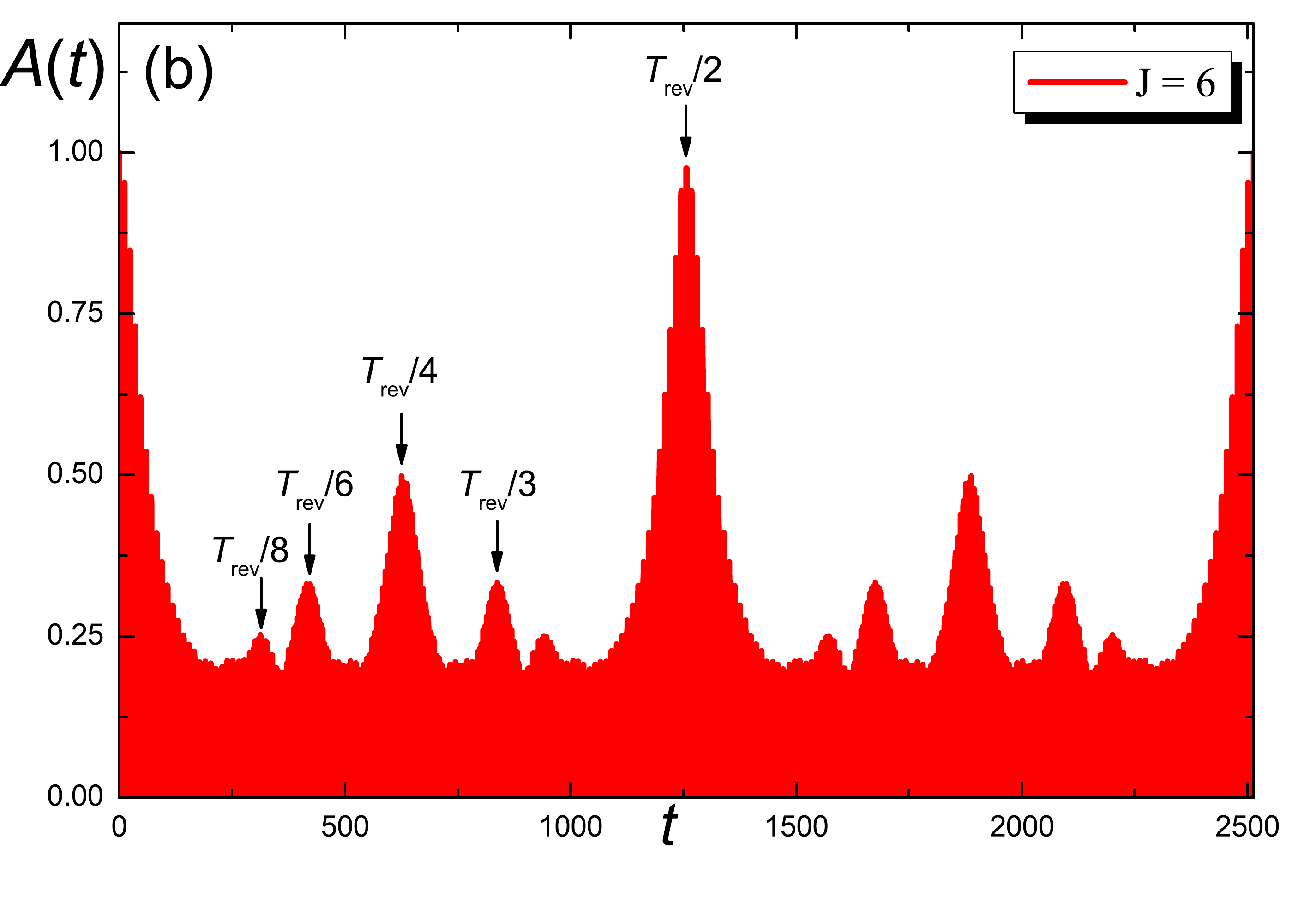}
\caption{(a) Autocorrelation function as a function of time for $J=1.5$, $
\protect\tau =0.1$, $\protect\omega =0.5$, $\hbar =1$, $\protect\gamma =0$, $
T_{\text{cl}}=10.05$ and $T_{\text{rev}}=251.32$; (b) Autocorrelation
function as a function of time for $J=6$, $\protect\tau =0.01$, $\protect
\omega =0.5$, $\hbar =1$, $\protect\gamma =0$, $T_{\text{cl}}=11.74$ and $T_{
\text{rev}}=2513.27$ \cite{Dey_Fring_squeezed}.}
\label{figGK1}
\end{figure*}
with $J\in \mathbb{R}_{0}^{+},\gamma \in \mathbb{R}$. In order to test
the quality of the coherent states, i.e. to see how closely they resemble
classical mechanics, we test Ehrenfest's theorem for an operator $\mathcal{A}$
\begin{eqnarray}
&& i\hbar \frac{d}{dt}\left\langle J,\gamma +t\omega ,\Phi \right\vert
\mathcal{A}\left\vert J,\gamma +t\omega ,\Phi \right\rangle _{\eta } \label{Ehren} \\
&& ~~~~ =\left\langle
J,\gamma +t\omega ,\Phi \right\vert [\mathcal{A},H]\left\vert J,\gamma +t\omega ,\Phi
\right\rangle _{\eta }. \notag
\end{eqnarray}
We used in \myref{Ehren} the fact that the time evolution for the states $
\left\vert J,\gamma ,\Phi \right\rangle $ is simply implemented as $\exp
(-iHt/\hbar )\left\vert J,\gamma ,\Phi \right\rangle =\left\vert J,\gamma
+t\omega ,\Phi \right\rangle $; see, \cite{Gazeau_Klauder,Antoine_Gazeau_Monceau_Klauder_Penson}. By computing the expectation values of the position and momentum observables and their squares for the NC non-Hermitian system, we obtain \cite{Dey_Fring_squeezed}
\begin{eqnarray}
\Delta X\Delta P &=& \frac{\hbar }{2}\left[ 1+\frac{\tau }{2}\left( 1+4J\sin
^{2}\gamma \right) \right] \\
&=& \frac{\hbar }{2}\left( 1+\hat{\tau}\left\langle
J,\gamma ,\Phi \right\vert P^{2}\left\vert J,\gamma ,\Phi \right\rangle
\right), \notag
\end{eqnarray}
where $\hat{\tau}=\tau\sqrt{\hbar/m\omega}$. This means that in the non-Hermitian setting the minimal uncertainty
product for the observables $X$ and $P$ is saturated and, the GK-coherent states $\left\vert J,\gamma ,\Phi \right\rangle $ are intelligent states. Remarkably this holds irrespective of the values for $J$ and $\gamma$. Next we verify Ehrenfest's theorem \myref{Ehren} for the operators $X$ and $P$ as well as the Newton's equation of motion; see, \cite{Dey_Fring_squeezed} for more details. The revival structure can be studied for the system by computing the autocorrelation function \cite{Dey_Fring_squeezed}
\begin{equation}
A(t):=\left\vert \left\langle J,\gamma ,\phi \right. \left\vert J,\gamma
+t\omega ,\phi \right\rangle \right\vert ^{2}=\left\vert \left\langle
J,\gamma ,\Phi \right. \left\vert J,\gamma +t\omega ,\Phi \right\rangle
_{\eta }\right\vert ^{2}.
\end{equation}
which is analyzed in Fig. \ref{figGK1}. In panel (a) of Fig. \ref{figGK1} we clearly observe local maxima at multiples of the classical period $T_{\text{cl}}$. As explained in \cite{Averbukh} the first full reconstruction of the original wave packet is obtained at $T_{\text{rev}}/2$ which is clearly visible in panel (a). The fractional revivals are better observed for smaller values of $\tau $ as depicted in panel (b). In that scenario the classical periods are so small as compared to the revival time that they are no longer resolved. We clearly observe a number of fractional revivals \cite{Dey_Fring_squeezed}. 

We also studied the q-deformed GK-coherent states for our system \cite{Dey_Fring_Gouba_Castro}. The predominant features of such states are that unlike the case discussed above, we do not obtain an intelligent states for all values of time, but for $t=0$, which is already an indication that the state is nonclassical in nature. This is more obvious while we study the revival properties of such states. We obtain a fractional superrevival structure and, thus, the $q$-deformed GK-coherent states are more nonclassical than the ordinary GK-coherent states.
\section{\textbf{Nonclassical states for non-Hermitian systems}}\label{sec5} 
\subsection{Squeezed states}
Squeezed states are one of the most important nonclassical states. Squeezed states are obtained by applying the Glauber's unitary displacement operator $D(\alpha$) on the squeezed vacuum \cite{Nieto_Truax_PRL}
\begin{figure*}
\centering   \includegraphics[width=9.0cm,height=6.2cm]{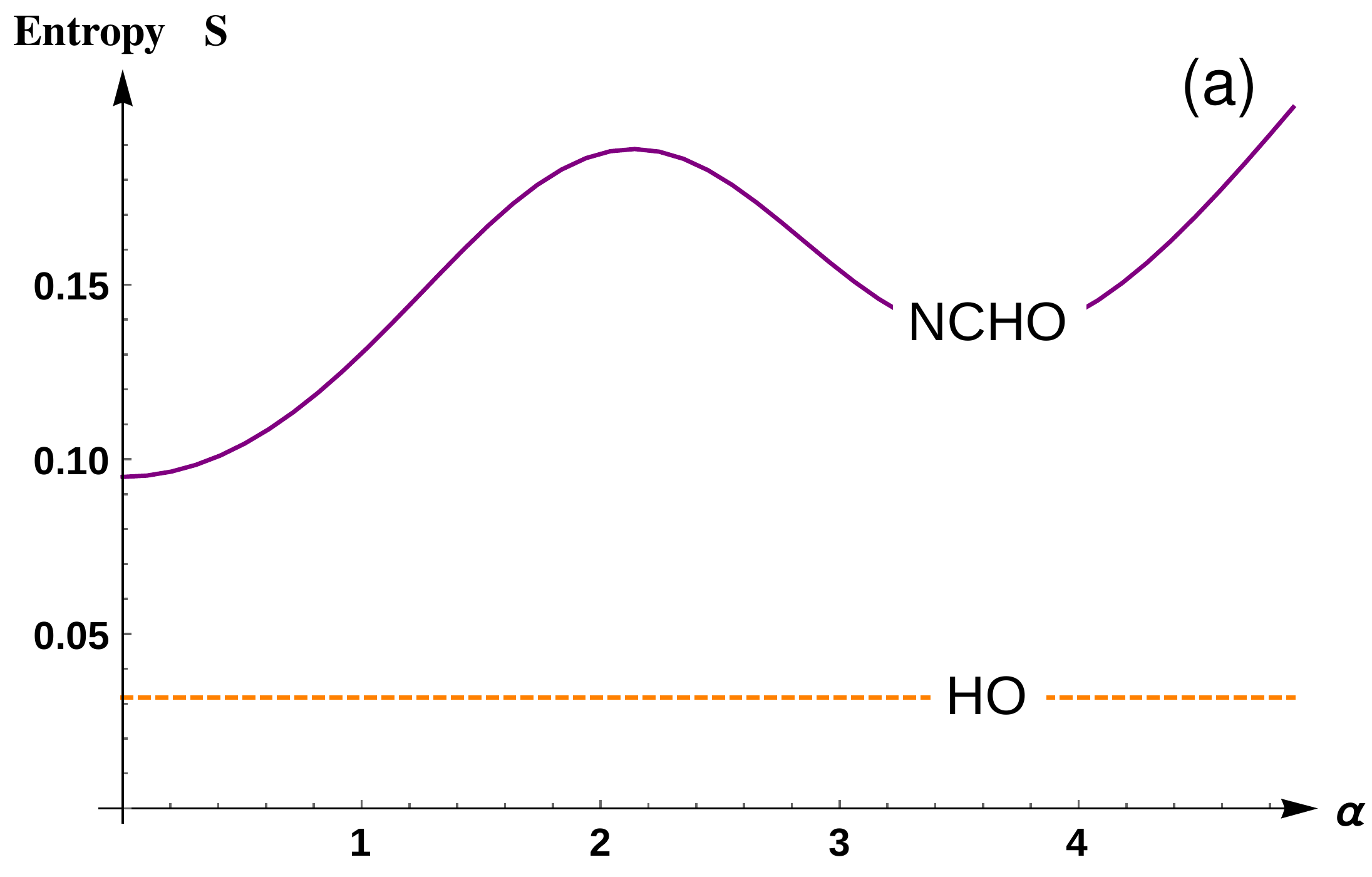}
\includegraphics[width=9.0cm,height=6.2cm]{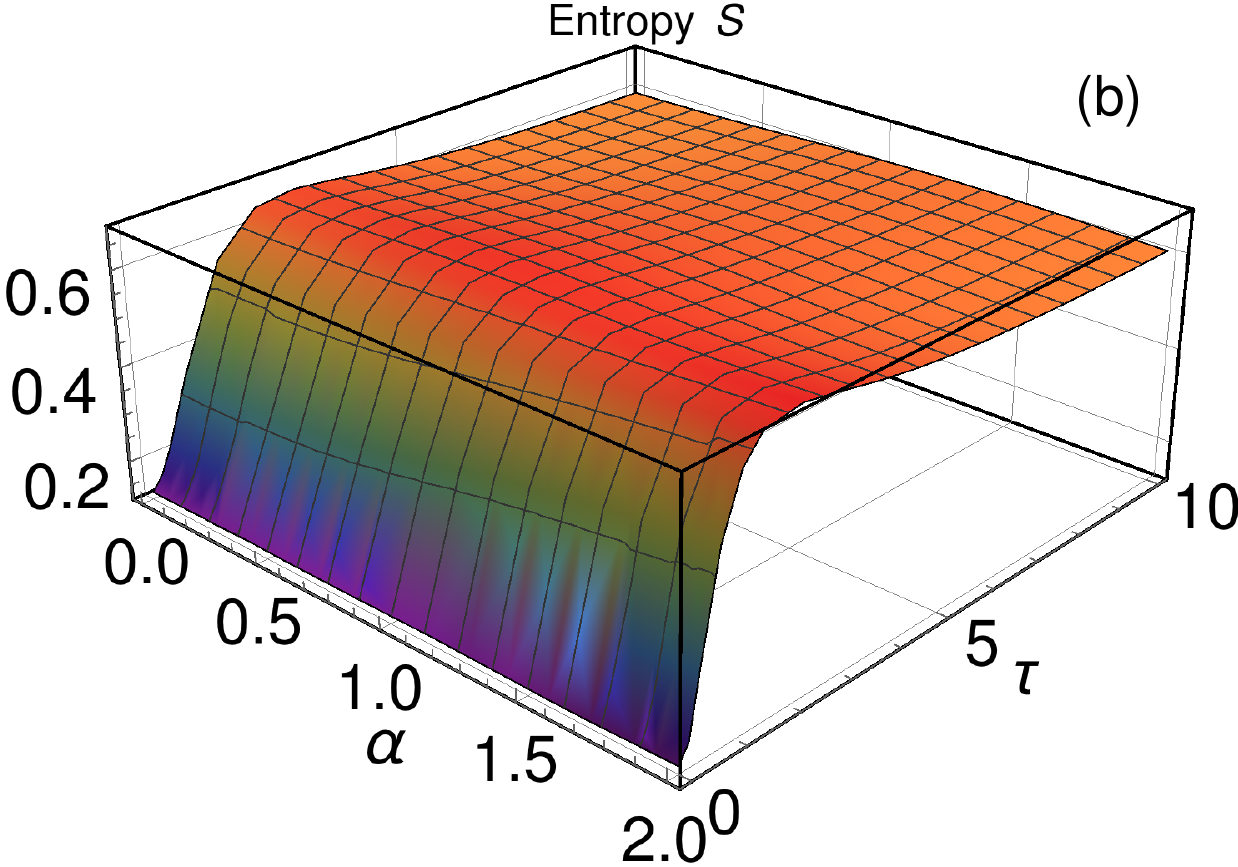}
\caption{\small{(a) Linear entropy for the NC squeezed state (solid, purple) versus squeezed state of ordinary harmonic oscillator (dashed, orange) as function of $\alpha$ for $\tau=0.5,\zeta=0.25$. Number of energy levels considered = 40 in each case. (b) Linear entropy for the NC squeezed state input as functions of $\alpha$ and $\tau$ for $\zeta=0.5$. Number of energy levels considered = 10 \cite{Dey_Hussin}.}}
\label{fig2}
\end{figure*}
\begin{eqnarray}\label{SqOperator}
\vert \alpha,\zeta\rangle &=& D(\alpha)S(\zeta)\vert 0\rangle, \quad S(\zeta)=e^{\frac{1}{2}(\zeta a^\dagger a^\dagger-\zeta^\ast aa)},\notag \\
D(\alpha) &=& e^{\alpha a^\dagger-\alpha^\ast a}, \quad \alpha, \zeta \in \mathbb{C},
\end{eqnarray}
with $\alpha$, $\zeta$ being displacement and squeezing parameters, respectively, and $S(\zeta)$ being the unitary squeezing operator. The ordering of $D(\alpha)S(\zeta)$ and $S(\zeta)D(\alpha)$ in \myref{SqOperator} are equivalent, amounting to a change of parameter \cite{Nieto_Truax_PRL}. An alternative ladder operator definition of the squeezed states can be obtained by performing the Holstein-Primakoff / Bogoliubov transformation on the squeezing operator \cite{Nieto_Truax_PRL}. The squeezed states $\vert \alpha,\zeta\rangle$ can be constructed from the solution of the equivalent ladder operator definition as follows \cite{Fu_Sasaki}
\begin{equation}\label{eigenSq}
(a+\zeta a^\dagger) \vert \alpha,\zeta\rangle=\alpha \vert \alpha,\zeta\rangle, \qquad \alpha,\zeta \in \mathbb{C}~.
\end{equation}
The coherent states are the special solutions when $\zeta=0$. A direct generalization \cite{Angelova_Hertz_Hussin} of the above definition is carried out by replacing the bosonic creation and annihilation operators $a, a^\dagger$ by the nonlinear ladder operators $A, A^\dagger$ \myref{genladder}. In order to solve the eigenvalue equation \myref{eigenSq} for the generalized case, let us expand the squeezed states $\vert \alpha,\zeta\rangle$ in terms of Fock states
\begin{equation}\label{Expansion}
\vert \alpha,\zeta\rangle=\frac{1}{\mathcal{N}(\alpha,\zeta)}\displaystyle\sum_{n=0}^\infty \frac{\mathcal{I}(\alpha,\zeta,n)}{\sqrt{n!}f(n)!}\vert n \rangle~.
\end{equation}
Inserting \myref{Expansion} into the eigenvalue equation \myref{eigenSq} replaced with the generalized ladder operators \myref{genladder}, we end up with a three terms recurrence relation
\begin{equation}\label{recurrence}
\mathcal{I}(\alpha,\zeta,n+1)=\alpha~\mathcal{I}(\alpha,\zeta,n)-\zeta nf^2(n) \mathcal{I}(\alpha,\zeta,n-1),
\end{equation}
with $\mathcal{I}(\alpha,\zeta,0)=1$ and $\mathcal{I}(\alpha,\zeta,1)=\alpha$, which when solved, leads to the explicit form of the squeezed states for the models corresponding to the particular values of $f(n)$ \cite{Angelova_Hertz_Hussin}. Note that, the recurrence relation \myref{recurrence} may not be easy to solve, when one deals with the complicated choices of $f(n)$ as in \myref{fn}. In our case, the solution is obtained in terms of the Gauss hypergeometric function $_2F_1$ as follows \cite{Dey_Hussin}
\begin{eqnarray}\label{SqueezedState}
\mathcal{I}(\alpha,&&\zeta,n)=i^n\left(\zeta B\right)^{n/2}\left(1+\frac{A}{B}\right)^{(n)} \\
&& \times 2F_1\Bigg[-n,\frac{1}{2}+\frac{A}{2B}+\frac{i\alpha}{2\sqrt{\zeta B}};1+\frac{A}{B};2\Bigg]~, \notag
\end{eqnarray}
so that by following the similar logic as given for NLCS in \myref{NcCoherent}-\myref{DisplacedCoh}, we obtain the explicit form of the squeezed states for our system \cite{Dey_Hussin}
\begin{equation}\label{SqStateNC}
\vert \alpha,\zeta\rangle =\frac{1}{\mathcal{N}(\alpha,\zeta)}\displaystyle\sum_{n=0}^{\infty} \frac{\mathcal{S}(\alpha,\zeta,n)}{\sqrt{n!}f(n)!}\vert n\rangle~,
\end{equation}
with
\begin{eqnarray}
&&\mathcal{S}(\alpha,\zeta,n)= \\
&&\left\{ \begin{array}{ll}
\mathcal{I}(\alpha,\zeta,n)-\frac{\tau}{16}\frac{f(n)!}{f(n+4)!}\mathcal{I}(\alpha,\zeta,n+4),~~~~0\leq n \leq 3
& \\ \mathcal{I}(\alpha,\zeta,n)-\frac{\tau}{16}\frac{f(n)!}{f(n+4)!}\mathcal{I}(\alpha,\zeta,n+4) & \\
~~~~~~~~~~~~+\frac{\tau}{16}\frac{n!}{(n-4)!}\frac{f(n)!}{f(n-4)!}\mathcal{I}(\alpha,\zeta,n-4),~~~n\geq 4. & \end{array}\right. \notag
\end{eqnarray}
In the harmonic oscillator limit $\tau=0$, i.e. $f(n)=1$, our expression for squeezed states reduces to that of the ordinary harmonic oscillator precisely
\begin{equation}\label{SqStateHO}
\vert \alpha,\zeta\rangle_{ho}^{}=\frac{1}{\mathcal{N}(\alpha,\zeta)}\displaystyle\sum_{n=0}^\infty \frac{1}{\sqrt{n!}}\left(\frac{\zeta}{2}\right)^{n/2}\mathcal{H}_n(\frac{\alpha}{\sqrt{2\zeta}})\vert n \rangle~,
\end{equation}
where $\mathcal{H}_n(\alpha)$ denote the Hermite polynomials. What we are left with is to quantify the nonclassicality that our states possess. For this, we consider the description given in Sec. \ref{SubSec25} and, thereafter, follow similar steps as Sec. \ref{SubBeam} to compute the beam splitter quantum entanglement. Here, we consider the squeezed state \myref{SqStateNC} at one of the inputs, while a vacuum state at the other.  We study a mutual comparison of the entanglement of the NC squeezed states \myref{SqStateNC} input with that of the squeezed state of the ordinary harmonic oscillators \myref{SqStateHO}. The outcomes for different values of the squeezing parameters have been demonstrated in panel (a) of Fig. \ref{fig2}. The linear entropies of the NC oscillators are much higher than the usual ones for all values of $\alpha$, which indicate that the output states resulting from the squeezed states of the NCHO are more entangled than that of the squeezed states of ordinary harmonic oscillators. The most exciting effect is that the key role on the behavior of the linear entropy is played by the NC parameter $\tau$, which is quite obvious in panel (b) of Fig. \ref{fig2}. The value of the entropy for the NC case coincides with the entropy of the ordinary harmonic oscillator, when $\tau=0$, which is expected. However, the entropy increases rapidly with the increase of the value of $\tau$ and saturates at sufficiently high value, irrespective of all values of $\alpha$ \cite{Dey_Hussin}. 
\subsection{Schr\"odinger cat states}
Cat states are another interesting type of nonclassical states that result from the superposition of two coherent states. The $q$-deformed version of cat states corresponding to our system reads as \cite{Dey}
\begin{eqnarray}
\big\vert\alpha\big\rangle_{q,\pm} &=& \frac{1}{\mathcal{N}_{q,\pm}(\alpha)}\Big(\vert\alpha\rangle_q\pm\vert-\alpha\rangle_q\Big),
\end{eqnarray} 
with the normalization constant
\begin{eqnarray}
\mathcal{N}_{q,\pm}^2(\alpha) &=& 2\pm\frac{2}{\mathcal{N}_{q,\pm}^2(\alpha)}\displaystyle\sum_{n=0}^{\infty}\frac{(-1)^n\vert\alpha\vert^{2n}}{[n]_q!} \notag\\
&=&~2\Big\{1\pm E_q(-2\vert\alpha\vert^2)\Big\},
\end{eqnarray}
which are, sometimes, also familiar as even and odd coherent states \cite{Xia_Guo}. A detailed analysis of the nonclassical properties of the given system can be found in \cite{Dey}, where we not only obtain the higher degree of nonclassicality of our system in comparison to the cat states of the harmonic oscillator, but also we notice that a suitable choice of parameter leads our system to the ideal squeezed states. A similar type of conclusion also emerges in the case of the cat states for the perturbative NCHO, which we studied in \cite{Dey_Fring_Hussin}.
\subsection{Photon-added coherent states}
The PACS \cite{Agarwal_Tara} are yet another interesting class of nonclassical states, which are obtained by $m$ successive actions of the canonical creation operator $a^\dagger$ on the standard coherent state $\vert\alpha\rangle$, as given by
\begin{equation}\label{PhotnAdded}
\vert\alpha,m\rangle =\frac{1}{\mathcal{N}(\alpha,m)}a^{\dagger m}\vert\alpha\rangle,
\end{equation}
with the normalization constant being $\mathcal{N}^2(\alpha,m) = \langle\alpha\vert a^ma^{\dagger m}\vert\alpha\rangle$. The PACS \myref{PhotnAdded} have been studied by many authors in different contexts both theoretically  \cite{Sivakumar_PACS,Duc_Noh} and experimentally \cite{Zavatta_Viciani_Bellini}. Nevertheless, for our systems the explicit expression of the PACS turns out to be \cite{Dey_Hussin_2}
\begin{eqnarray}\label{qPACS}
\vert\alpha,m\rangle_q &=& \frac{1}{\mathcal{N}_q(\alpha,m)}A_q^{\dagger m}\vert\alpha\rangle_q \\
&=& \frac{1}{\mathcal{N}_q(\alpha,m)\mathcal{N}_q(\alpha)}\displaystyle\sum_{n=0}^{\infty}\frac{\alpha^n}{[n]_q!}\sqrt{[n+m]_q!}~\vert n+m\rangle_q, \notag
\end{eqnarray}
with the normalization constant
\begin{eqnarray}
\mathcal{N}_q^2(\alpha,m) &=& {}_q\langle\alpha,m\vert A_q^mA_q^{\dagger m}\vert\alpha,m\rangle_q \\
&=& \frac{1}{\mathcal{N}_q^2(\alpha)}\displaystyle\sum_{n=0}^{\infty}\frac{\vert\alpha\vert^{2n}}{[n]_q!^2}[n+m]_q!, \notag
\end{eqnarray}
where $\vert\alpha\rangle_q$ is a standard $q$-deformed coherent state \myref{qdefcoherent} and $\mathcal{N}_q^2(\alpha)$ is the corresponding normalization constant \myref{NormqD}. In order to analyze the nonclassical properties of the deformed PACS \myref{qPACS}, we studied two types of higher-order squeezing properties of the quadratures; namely, Hillery-type \cite{Hillery} and Hong--Mandel-type \cite{Hong_Mandel}. Both of the studies lead to the overall conclusion that irrespective of the the order of squeezing $N$ and number of photons added $m$, deformed PACS are always more nonclassical than those of the ordinary harmonic oscillator. In addition, by increasing the deformation of the system, it is possible to increase the nonclassicality of the system accordingly. The phenomena is also supported by the analysis of the higher-order photon number squeezing in terms of the study of the Mandel parameter and second order correlation function both in higher orders \cite{Duc_Noh}. For detailed mathematical analysis on these; see, \cite{Dey_Hussin_2}, where we encountered many difficulties to handle such problems for the given mathematical structure. It should also be mentioned that all of our analysis is nontrivial and it has been carried out in a complete generic way so that the method can be applied to any other deformed systems.
\section{\textbf{Applications}}\label{sec6} 
Some crucial applications of the coherent states for non-Hermitian system have been studied in various different contexts. They have been applied to the study of mean-field dynamics of Bose-Einstein condensation \cite{Graefe_PRL}, in the construction of coherent states for time-dependent systems \cite{Dey_Fring_Time}, in the study of quantum tomogram \cite{Jayakrishnan_Dey}, in quantum evolution \cite{Graefe_Schubert}, in the study of Fermionic coherent states \cite{Cherbal}, $\lambda$ coherent states \cite{Beckers}, bicoherent states \cite{Bagarello_BiCoherent}, path integral coherent states \cite{Kandirmaz}, complex oscillator systems \cite{Zelaya2017}, systems with position dependent mass \cite{Yahiaoui} etc. Moreover, as already discussed that our systems being well-connected with the nonlinear coherent states, they can be applied to anywhere where the usual nonlinear coherent states are applied. Nevertheless, throughout the article it is obvious that the theoretical part of the study of coherent and  nonclassical states for non-Hermitian system has been impressive, since it provides significant advancements in the subjects of quantum optics and information theory. However, what is more important is to understand the process of creating such systems in the laboratory. For this, we do not have a clear picture yet, however, the good news is that our systems have a good connection with nonlinear coherent states. So, if the nonlinear coherent states can be found to exist in real life experiment, the study of our system in the laboratory will not be far. Experimental attempts for nonlinear coherent states have been enormous \cite{Wang_Goorskey_Xiao,Gambetta2006,Yan_Zhu_Li}, for more references; see, \cite{Dodonov_Review}. 
\section{\textbf{Concluding remarks}}\label{sec7} 
We have studied several different type of coherent and nonclassical states for non-Hermitian models emerging from an interesting NC framework. Specifically, we observe that the NLCS discussed in Sec. \ref{SubSec41} possess nonclassical properties, in fact, they behave as ideal squeezed states. Therefore, although the states emerged from a coherent state definition, they do not have any classical analogue. In turn, the $q$-deformed coherent states investigated in Sec. \ref{SubSec42} characteristically can be attributed to good coherent states. Indeed, they are intelligent states, thus, qualitatively they are equivalent to the famous Glauber coherent states. However, by looking at the sub-Poissonian distribution, we notice that the state possesses a slight nonclassicality. A similar behavior is observed for the GK-coherent states also that has been analyzed in Sec. \ref{SubSec43}. They are intelligent, however, they exhibit fractional revival structure. The $q$-deformed version of GK-coherent states are even more nonclassical, since they show fractional super-revival structure.

We also have several interesting observations in Sec. \ref{sec5} during the study of various nonclassical states; such as, squeezed states, cat states, photon-added coherent states. All of the studies support that a larger deformation parameter leads to higher degree of nonclassicality. The deformation parameter being the characteristics of the models itself, the higher degree of nonclassciality for any noncalsscical states is inherited by the models. Moreover, we have developed some concepts that help to quantify the amount of nonclassicality that a state possess and, thus, exploring new opportunities towards the quantum information theory. Thus, our work connects a bigger mathematical structure with the quantum optical systems leading to interesting possibilities of several new directions of advancements of the theory.

\vspace{0.5cm} \noindent \textbf{Acknowledgements:} S.D. is supported by an INSPIRE Faculty Grant (DST/INSPIRE/04/2016/001391) by the Department of Science and Technology, Government of India. V.H. acknowledges the support of research grants from NSERC of Canada


\begin{thebibliography}{300}

\bibitem{Schrodinger}
E.~Schr{\"o}dinger,
\newblock Der stetige {\"u}bergang von der mikro-zur makromechanik,
\newblock {Naturwissenschaften} \textbf{14}, 664--666 (1926).

\bibitem{Kennard}
E.~H. Kennard,
\newblock Zur quantenmechanik einfacher bewegungstypen,
\newblock {Z. Phys.} \textbf{44}, 326--352 (1927).

\bibitem{Darwin}
C.~G. Darwin,
\newblock Free motion in the wave mechanics,
\newblock {Proc. Royal Soc. London} \textbf{117}, 258--293 (1927).

\bibitem{Glauber}
R.~J. Glauber,
\newblock Coherent and incoherent states of the radiation field,
\newblock {Phys. Rev.} \textbf{131}, 2766 (1963).

\bibitem{Barut_Girardello}
A.~O. Barut and L.~Girardello,
\newblock New ``coherent” states associated with non-compact groups,
\newblock {Commun. Math. Phys.} \textbf{21}, 41--55 (1971).

\bibitem{Perelomov}
A.~M. Perelomov,
\newblock Coherent states for arbitrary {Lie} group,
\newblock {Commun. Math. Phys.} \textbf{26}, 222--236 (1972).

\bibitem{Gilmore}
R.~Gilmore,
\newblock Geometry of symmetrized states,
\newblock {Ann. Phys.} \textbf{74}, 391--463 (1972).

\bibitem{Gazeau_Klauder}
J.~P. Gazeau and J.~R. Klauder,
\newblock Coherent states for systems with discrete and continuous spectrum,
\newblock {J. Phys. A: Math. Gen.} \textbf{32}, 123 (1999).

\bibitem{Ali_Antoine_Gazeau}
S.~T. Ali, J.~P. Antoine and J.~P. Gazeau,
\newblock {\em Coherent states, wavelets and their generalizations},
\newblock Springer-Verlag: New York (2000).

\bibitem{Bahr_Thiemann}
B.~Bahr and T.~Thiemann,
\newblock Gauge-invariant coherent states for loop quantum gravity: II.
  {Non-Abelian} gauge groups,
\newblock {Class. Quantum Grav.} \textbf{26}, 045012 (2009).

\bibitem{Hawking}
S.~W. Hawking,
\newblock Quantum coherence down the wormhole,
\newblock {Phys. Lett. B} \textbf{195}, 337--343 (1987).

\bibitem{Gazeau_Book}
J.-P. Gazeau,
\newblock {\em Coherent states in quantum physics},
\newblock Wiley-VCH: Berlin (2009).

\bibitem{Kim_Son_Buzek_Knight}
M.~S. Kim, W.~Son, V.~Bu{\v{z}}ek and P.~L. Knight,
\newblock Entanglement by a beam splitter: {Nonclassicality} as a prerequisite for entanglement,
\newblock {Phys. Rev. A} \textbf{65}, 032323 (2002).

\bibitem{Klauder_Skagerstam_Book}
J.~R. Klauder and B.-S. Skagerstam,
\newblock {\em Coherent states: applications in physics and mathematical
  physics},
\newblock World scientific: Singapore (1985).

\bibitem{Perelomov_Book}
A.~Perelomov,
\newblock {\em Generalized coherent states and their applications},
\newblock Springer-Verlag: Berlin (1986).

\bibitem{Gerry_Knight_Book}
C.~Gerry and P.~Knight,
\newblock {\em Introductory quantum optics},
\newblock Cambridge Univ. Press: New York (2005).

\bibitem{Combescure_Robert_Book}
M.~Combescure and D.~Robert,
\newblock {\em Coherent states and applications in mathematical physics},
\newblock Springer (2012).

\bibitem{Zhang_Review}
W.-M. Zhang, D.~H. Feng and R.~Gilmore,
\newblock Coherent states: theory and some applications,
\newblock {Rev. Mod. Phys.} \textbf{62}, 867 (1990).

\bibitem{Dodonov_Review}
V.~V. Dodonov,
\newblock Nonclassical states in quantum optics: a squeezed review of the first 75 years,
\newblock {J. Opt. B: Quant. Semiclas. Opt.} \textbf{4}, R1 (2002).

\bibitem{Dodonov_Manko_Book}
V.~V. Dodonov and V.~I. Man'ko,
\newblock {\em Theory of nonclassical states of light},
\newblock CRC Press: London (2003).

\bibitem{Sanders_Review}
B.~C. Sanders,
\newblock Review of entangled coherent states,
\newblock {J. Phys. A: Math. Theor.} \textbf{45}, 244002 (2012).

\bibitem{Dey_Fring_squeezed}
S.~Dey and A.~Fring,
\newblock Squeezed coherent states for noncommutative spaces with minimal
  length uncertainty relations,
\newblock {Phys. Rev. D} \textbf{86}, 064038 (2012).

\bibitem{Dey_Fring_Gouba_Castro}
S.~Dey, A.~Fring, L.~Gouba and P.~G. Castro,
\newblock Time-dependent $q$-deformed coherent states for generalized
  uncertainty relations,
\newblock {Phys. Rev. D} \textbf{87}, 084033 (2013).

\bibitem{Dey}
S.~Dey,
\newblock $q$-deformed noncommutative cat states and their nonclassical
  properties,
\newblock {Phys. Rev. D} \textbf{91}, 044024 (2015).

\bibitem{Dey_Hussin}
S.~Dey and V.~Hussin,
\newblock Entangled squeezed states in noncommutative spaces with minimal
  length uncertainty relations,
\newblock {Phys. Rev. D} \textbf{91}, 124017 (2015).

\bibitem{Dey_Hussin_2}
S.~Dey and V.~Hussin,
\newblock Noncommutative $q$-photon-added coherent states,
\newblock {Phys. Rev. A} \textbf{93}, 053824 (2016).

\bibitem{Dey2}
S.~Dey,
\newblock On completeness of coherent states in noncommutative spaces with
  generalised uncertainty principle,
\newblock {arXiv:1609.04460}.

\bibitem{Dey_Fring_Hussin}
S.~Dey, A.~Fring and V.~Hussin,
\newblock Nonclassicality versus entanglement in a noncommutative space,
\newblock {Int. J. Mod. Phys. B} \textbf{31}, 1650248 (2017).

\bibitem{Stoler}
D.~Stoler,
\newblock Generalized coherent states,
\newblock {Phys. Rev. D} \textbf{4}, 2309 (1971).

\bibitem{Arik_Coon}
M.~Arik and D.~D. Coon,
\newblock Hilbert spaces of analytic functions and generalized coherent states,
\newblock {J. Math. Phys.} \textbf{17}, 524--527 (1976).

\bibitem{Manko_Marmo_Sudarshan_Zaccaria}
V.~I. Man'ko, G.~Marmo, E.~C.~G. Sudarshan and F.~Zaccaria,
\newblock $f$-oscillators and nonlinear coherent states,
\newblock {Phys. Scr.} \textbf{55}, 528 (1997).

\bibitem{Sivakumar}
S.~Sivakumar,
\newblock Studies on nonlinear coherent states,
\newblock {J. Opt. B: Quant. Semiclas. Opt.} \textbf{2}, R61 (2000).

\bibitem{Fox_Choi}
R.~F. Fox and M.~H. Choi,
\newblock Generalized coherent states and quantum-classical correspondence,
\newblock {Phys. Rev. A} \textbf{61}, 032107 (2000).

\bibitem{Dey_Fring_PRA}
S.~Dey and A.~Fring,
\newblock Bohmian quantum trajectories from coherent states,
\newblock {Phys. Rev. A} \textbf{88}, 022116 (2013).

\bibitem{iwata}
G.~Iwata,
\newblock {Non-Hermitian operators and eigenfunction expansions},
\newblock {Prog. Theor. Phys.} \textbf{6}, 216--226 (1951).

\bibitem{Loudon_Knight}
R.~Loudon and P.~L. Knight,
\newblock Squeezed light,
\newblock {J. Mod. Opt.} \textbf{34}, 709--759 (1987).

\bibitem{Antoine_Gazeau_Monceau_Klauder_Penson}
J.-P. Antoine~et al.,
\newblock Temporally stable coherent states for infinite well and
  {P{\"o}schl--Teller} potentials,
\newblock {J. Math. Phys.} \textbf{42}, 2349--2387 (2001).

\bibitem{Aragone_Guerri_Salamo_Tani}
C.~Aragone, G.~Guerri, S.~Salamo and J.~L. Tani,
\newblock Intelligent spin states,
\newblock {J. Phys. A: Math. Nucl. Gen.} \textbf{7}, L149 (1974).

\bibitem{Trifonov5}
D. A. ~Trifonov,
\newblock Generalized intelligent states and squeezing,
\newblock {J. Math. Phys.} \textbf{35}, 2297--2308 (1994).

\bibitem{Trifonov6}
D. A. ~Trifonov,
\newblock Robertson intelligent states,
\newblock {J. Phys. A: Math. Gen.} \textbf{30}, 5941 (1997).

\bibitem{Glauber3}
R.~J Glauber,
\newblock Photon correlations,
\newblock {Phys. Rev. Lett.} \textbf{10}, 84 (1963).

\bibitem{Sudarshan2}
E.~C.~G. Sudarshan,
\newblock Equivalence of semiclassical and quantum mechanical descriptions of
  statistical light beams,
\newblock {Phys. Rev. Lett.} \textbf{10}, 277--279 (1963).

\bibitem{Husimi}
K.~Husimi,
\newblock Some formal properties of the density matrix,
\newblock {Proc. Phys. Math. Soc. Jpn.} \textbf{22}, 264--314 (1940).

\bibitem{Wigner}
E.~Wigner,
\newblock On the quantum correction for thermodynamic equilibrium,
\newblock {Phys. Rev.} \textbf{40}, 749 (1932).

\bibitem{Mandel_Wolf_Book}
L.~Mandel and E.~Wolf,
\newblock {\em Optical coherence and quantum optics},
\newblock Cambridge Univ. Press: New York (1995).

\bibitem{Scully_Zubairy_Book}
M.~O. Scully and M.~S. Zubairy,
\newblock {\em Quantum optics},
\newblock Cambridge Univ. Press: UK (1997).

\bibitem{Klauder_Sudarshan_Book}
J.~R. Klauder and E.~C.~G. Sudarshan,
\newblock {\em Fundamentals of quantum optics},
\newblock Dover Pub.: New York (2006).

\bibitem{Walls_Milburn_Book}
D.~F. Walls and G.~J. Milburn,
\newblock {\em Quantum optics},
\newblock Springer (2007).

\bibitem{Agarwal_Book}
G.~S. Agarwal,
\newblock {\em Quantum optics},
\newblock Cambridge Univ. Press: UK (2013).

\bibitem{Johansen}
L.~M. Johansen,
\newblock Nonclassical properties of coherent states,
\newblock {Phys. Lett. A} \textbf{329}, 184--187 (2004).

\bibitem{Helstrom}
C.~Helstrom,
\newblock Nonclassical states in optical communication to a remote receiver
  (Corresp.),
\newblock {IEEE Trans. Inf. Theory} \textbf{26}, 378--382 (1980).

\bibitem{Lugiato}
L.~A. Lugiato and G.~Strini,
\newblock On nonclassical effects in two-photon optical bistability and
  two-photon laser,
\newblock {Opt. Commun.} \textbf{41}, 374--378 (1982).

\bibitem{Mollow_Glauber}
B.~R. Mollow and R.~J. Glauber,
\newblock Quantum theory of parametric amplification. I,
\newblock {Phys. Rev.} \textbf{160}, 1076 (1967).

\bibitem{Aharonov_Falkoff}
Y.~Aharonov, D.~Falkoff, E.~Lerner and H.~Pendleton,
\newblock A quantum characterization of classical radiation,
\newblock {Ann. Phys.} \textbf{39}, 498--512 (1966).

\bibitem{Hillery2}
M.~Hillery,
\newblock Classical pure states are coherent states,
\newblock {Phys. Lett. A} \textbf{111}, 409--411 (1985).

\bibitem{Titulaer_Glauber}
U.~M. Titulaer and R.~J. Glauber,
\newblock Correlation functions for coherent fields,
\newblock {Phys. Rev.} \textbf{140}, B676 (1965).

\bibitem{Gao}
W.~B. Gao~et al.,
\newblock Experimental demonstration of a hyper-entangled ten-qubit
  Schr{\"o}dinger cat state,
\newblock {Nature Phys.} \textbf{6}, 331--335 (2010).

\bibitem{Walls}
D.~F. Walls,
\newblock Squeezed states of light,
\newblock {Nature (London)} \textbf{306}, 141 (1983).

\bibitem{Agarwal_Tara}
G.~S. Agarwal and K.~Tara,
\newblock Nonclassical properties of states generated by the excitations on a
  coherent state,
\newblock {Phys. Rev. A} \textbf{43}, 492 (1991).

\bibitem{Xia_Guo}
Y.~Xia and G.~Guo,
\newblock Nonclassical properties of even and odd coherent states,
\newblock {Phys. Lett. A} \textbf{136}, 281--283 (1989).

\bibitem{Wakui}
K.~Wakui, H.~Takahashi, A.~Furusawa and M.~Sasaki,
\newblock Photon subtracted squeezed states generated with periodically poled
  {KTiOPO\textsubscript{4}},
\newblock {Opt. Exp.} \textbf{15}, 3568--3574 (2007).

\bibitem{Lee}
C.~T. Lee,
\newblock Measure of the nonclassicality of nonclassical states,
\newblock {Phys. Rev. A} \textbf{44}, R2775 (1991).

\bibitem{Marian}
P.~Marian, T.~A. Marian and H.~Scutaru,
\newblock Quantifying nonclassicality of one-mode Gaussian states of the
  radiation field,
\newblock {Phys. Rev. Lett.} \textbf{88}, 153601 (2002).

\bibitem{Caves}
C. M. Caves,
\newblock Quantum-mechanical noise in an interferometer,
\newblock {Phys. Rev. D} \textbf{23}, 1693 (1981).

\bibitem{Milburn_Walls5}
G. Milburn and D. F. Walls,
\newblock Production of squeezed states in a degenerate parametric amplifier,
\newblock {Opt. Commun.} \textbf{39}, 401--404 (1981).

\bibitem{Trifonov7}
D. A. Trifonov,
\newblock On the squeezed states for $n$ observables,
\newblock {Phys. Scr.} \textbf{58}, 246 (1998).

\bibitem{Mandel}
L.~Mandel,
\newblock {Sub-Poissonian} photon statistics in resonance fluorescence,
\newblock {Opt. Lett.} \textbf{4}, 205--207 (1979).

\bibitem{Glauber2}
R.~J. Glauber,
\newblock The quantum theory of optical coherence,
\newblock {Phys. Rev.} \textbf{130}, 2529 (1963).

\bibitem{Eberly}
J.~H. Eberly, N.~B. Narozhny and J.~J. Sanchez-Mondragon,
\newblock Periodic spontaneous collapse and revival in a simple quantum model,
\newblock {Phys. Rev. Lett.} \textbf{44}, 1323 (1980).

\bibitem{Averbukh}
I.~Sh. Averbukh and N.~F. Perelman,
\newblock Fractional revivals: Universality in the long-term evolution of
  quantum wave packets beyond the correspondence principle dynamics,
\newblock {Phys. Lett. A} \textbf{139}, 449--453 (1989).

\bibitem{Xiang-Bin}
W.~Xiang-Bin,
\newblock Theorem for the beam-splitter entangler,
\newblock {Phys. Rev. A} \textbf{66}, 024303 (2002).

\bibitem{Snyder}
H.~S. Snyder,
\newblock Quantized space-time,
\newblock {Phys. Rev.} \textbf{71}, 38 (1947).

\bibitem{Yang}
C.~N. Yang,
\newblock On quantized space-time,
\newblock {Phys. Rev.} \textbf{72}, 874 (1947).

\bibitem{Seiberg_Witten}
N.~Seiberg and E.~Witten,
\newblock String theory and noncommutative geometry,
\newblock {J. High Energy Phys.} \textbf{1999}, 032 (1999).

\bibitem{Connes}
A.~Connes,
\newblock Non-commutative differential geometry,
\newblock {Pub. Math. l'IH{\'E}S} \textbf{62}, 41--144 (1985).

\bibitem{Woronowicz}
S.~L. Woronowicz,
\newblock Compact matrix pseudogroups,
\newblock {Commun. Math. Phys.} \textbf{111}, 613--665 (1987).

\bibitem{Garay}
L.~J Garay,
\newblock Quantum gravity and minimum length,
\newblock {Int. J. Mod. Phys. A} \textbf{10}, 145--165 (1995).

\bibitem{Connes_Book}
A.~Connes,
\newblock {\em Noncommutative geometry},
\newblock Academic Press (1995).

\bibitem{Madore_Book}
J.~Madore,
\newblock {\em An introduction to noncommutative differential geometry and its physical applications},
\newblock Cambridge Univ. Press: UK (1999).

\bibitem{Douglas_Nekrasov}
M.~R. Douglas and N.~A. Nekrasov,
\newblock Noncommutative field theory,
\newblock {Rev. Mod. Phys.} \textbf{73}, 977 (2001).

\bibitem{Szabo}
R.~J. Szabo,
\newblock Quantum field theory on noncommutative spaces,
\newblock {Phys. Rep.} \textbf{378}, 207--299 (2003).

\bibitem{Doplicher}
S.~Doplicher, K.~Fredenhagen and J.~E. Roberts,
\newblock The quantum structure of spacetime at the Planck scale and quantum
  fields,
\newblock {Commun. Math. Phys.} \textbf{172}, 187--220 (1995).

\bibitem{Aschieri}
P.~Aschieri, M.~Dimitrijevi{\'c}, F.~Meyer and J.~Wess,
\newblock Noncommutative geometry and gravity,
\newblock {Class. Quantum Grav.} \textbf{23}, 1883 (2006).

\bibitem{Castro_Kullock_Toppan}
P.~G. Castro, R.~Kullock and F.~Toppan,
\newblock Snyder noncommutativity and pseudo-Hermitian Hamiltonians from a
  Jordanian twist.
\newblock {J. Math. Phys.} \textbf{52}, 062105 (2011).

\bibitem{Dey_Fring_Time}
S.~Dey and A.~Fring,
\newblock Noncommutative quantum mechanics in a time-dependent background,
\newblock {Phys. Rev. D} \textbf{90}, 084005 (2014).

\bibitem{Dey_Fring_Mathanaranjan}
S.~Dey, A.~Fring and T.~Mathanaranjan,
\newblock Spontaneous {PT-symmetry breaking for systems of noncommutative
  Euclidean Lie algebraic type},
\newblock {Int. J. Theor. Phys.} \textbf{54}, 4027--4033 (2015).

\bibitem{Gouba_Review}
L.~Gouba,
\newblock A comparative review of four formulations of noncommutative quantum
  mechanics,
\newblock {Int. J. Mod. Phys. A} \textbf{31}, 1630025 (2016).

\bibitem{Kempf_Mangano_Mann}
A.~Kempf, G.~Mangano and R.~B. Mann,
\newblock Hilbert space representation of the minimal length uncertainty
  relation,
\newblock {Phys. Rev. D} \textbf{52}, 1108 (1995).

\bibitem{Das_Vagenas}
S.~Das and E.~C. Vagenas,
\newblock Universality of quantum gravity corrections,
\newblock {Phys. Rev. Lett.} \textbf{101}, 221301 (2008).

\bibitem{Gomes_Kupriyanov}
M.~Gomes and V.~G. Kupriyanov,
\newblock Position-dependent noncommutativity in quantum mechanics,
\newblock {Phys. Rev. D} \textbf{79}, 125011 (2009).

\bibitem{Bagchi_Fring}
B.~Bagchi and A.~Fring,
\newblock Minimal length in quantum mechanics and {non-Hermitian Hamiltonian
  systems},
\newblock {Phys. Lett. A} \textbf{373}, 4307--4310 (2009).

\bibitem{Quesne_Tkachuk}
C.~Quesne and V.~M. Tkachuk,
\newblock Composite system in deformed space with minimal length,
\newblock {Phys. Rev. A} \textbf{81}, 012106 (2010).

\bibitem{Brau}
F.~Brau,
\newblock Minimal length uncertainty relation and the hydrogen atom,
\newblock {J. Phys. A: Math. Gen.} \textbf{32}, 7691 (1999).

\bibitem{Fring_Gouba_Scholtz}
A.~Fring, L.~Gouba and F.~G. Scholtz,
\newblock Strings from position-dependent noncommutativity,
\newblock {J. Phys. A: Math. Theor.} \textbf{43}, 345401 (2010).

\bibitem{Chang}
L.~N. Chang, Z.~Lewis, D.~Minic and T.~Takeuchi,
\newblock On the minimal length uncertainty relation and the foundations of
  string theory,
\newblock {Adv. High Energy Phys.} \textbf{2011}, 493514 (2011).

\bibitem{Nozari_Etemadi}
K.~Nozari and A.~Etemadi,
\newblock Minimal length, maximal momentum, and {Hilbert} space representation of quantum mechanics,
\newblock {Phys. Rev. D} \textbf{85}, 104029 (2012).

\bibitem{Maziashvili}
M.~Maziashvili,
\newblock Minimum-length deformed quantization of a free field on the de Sitter background and corrections to the inflaton perturbations,
\newblock {Phys. Rev. D} \textbf{85}, 125026 (2012).

\bibitem{Sprenger}
M.~Sprenger, P.~Nicolini and M.~Bleicher,
\newblock Physics on the smallest scales: an introduction to minimal length
  phenomenology,
\newblock {Euro. J. Phys.} \textbf{33}, 853 (2012).

\bibitem{Dey_Fring_Gouba}
S.~Dey, A.~Fring and L.~Gouba,
\newblock $\mathcal{PT}$-symmetric non-commutative spaces with minimal volume
  uncertainty relations,
\newblock {J. Phys. A: Math. Theor.} \textbf{45}, 385302 (2012).

\bibitem{Dey_Fring_Acta}
S.~Dey and A.~Fring,
\newblock The two-dimensional harmonic oscillator on a noncommutative space
  with minimal uncertainties,
\newblock {Acta Polytechnica} \textbf{53}, 268--270 (2013).

\bibitem{Dey_Fring_Khantoul}
S.~Dey, A.~Fring and B.~Khantoul,
\newblock Hermitian versus {non-Hermitian} representations for minimal length
  uncertainty relations,
\newblock {J. Phys. A: Math. Theor.} \textbf{46}, 335304 (2013).

\bibitem{Maziashvili_Megrelidze}
M.~Maziashvili and L.~Megrelidze,
\newblock Minimum-length deformed quantum mechanics/quantum field theory,
  issues, and problems,
\newblock {Prog. Theor. Exp. Phys.} \textbf{2013}, 123B06 (2013).

\bibitem{Dey_Thesis}
S.~Dey,
\newblock Solvable Models on Noncommutative Spaces with Minimal Length
  Uncertainty Relations,
\newblock {Ph. D. thesis (City, University of London, UK), arXiv:1410.3193}.

\bibitem{Sobhani_Hassanabadi}
H.~Sobhani and H.~Hassanabadi,
\newblock {Two-Dimensional Linear Dependencies on the Coordinate Time-Dependent Interaction in Relativistic Non-Commutative Phase Space},
\newblock {Commun. Theor. Phys.} \textbf{64}, 263 (2015).

\bibitem{Bhat_Dey_Faizal_Hou_Zhao}
A.~Bhat, S.~Dey, M.~Faizal, C.~Hou and Q.~Zhao,
\newblock Modification of {Schr{\"o}dinger--Newton} equation due to braneworld models with minimal length,
\newblock Phys. Lett. B \textbf{770}, 325–330 (2017).

\bibitem{Lewis_Roman_Takeuchi}
Z.~Lewis, A.~Roman and T.~Takeuchi,
\newblock Position and momentum uncertainties of a particle in a {V-shaped
  potential under the minimal length uncertainty relation},
\newblock {Int. J. Mod. Phys. A} \textbf{30}, 1550206 (2015).

\bibitem{Falaye}
B.~J. Falaye~et al.,
\newblock Massive fermions interacting via a harmonic oscillator in the
  presence of a minimal length uncertainty relation,
\newblock {Int. J. Mod. Phys. E} \textbf{24}, 1550087 (2015).

\bibitem{Nascimento_Aguiar_Guedes}
J.~P.~G. Nascimento, V.~Aguiar and I.~Guedes,
\newblock Entropy and information of a harmonic oscillator in a time-varying
  electric field in {2D and 3D} noncommutative spaces,
\newblock {Physica A} \textbf{477}, 65--77 (2017).

\bibitem{Mead}
C.~A. Mead,
\newblock Possible connection between gravitation and fundamental length,
\newblock {Phys. Rev.} \textbf{135}, B849 (1964).

\bibitem{Veneziano}
G.~Veneziano,
\newblock A stringy nature needs just two constants,
\newblock {Europhys. Lett.} \textbf{2}, 199 {1986}.

\bibitem{Rovelli}
C.~Rovelli,
\newblock Loop quantum gravity,
\newblock {Living Rev. Rel.} \textbf{1}, 1 (1998).

\bibitem{Padmanabhan}
T.~Padmanabhan,
\newblock Physical significance of {Planck} length,
\newblock {Ann. Phys.} \textbf{165}, 38--58 (1985).

\bibitem{Amelino}
G.~Amelino-Camelia,
\newblock Testable scenario for relativity with minimum length,
\newblock {Phys. Lett. B} \textbf{510}, 255--263 (2001).

\bibitem{Magueijo}
J.~Magueijo and L.~Smolin,
\newblock Lorentz invariance with an invariant energy scale,
\newblock {Phys. Rev. Lett.} \textbf{88}, 190403 (2002).

\bibitem{Trifonov_Review}
D.~A. Trifonov,
\newblock Generalized uncertainty relations and coherent and squeezed states,
\newblock {J. Opt. Soc. Am. A} \textbf{17}, 2486--2495 (2000).

\bibitem{Quesne_Penson_Tkachuk}
C.~Quesne, K.~A. Penson and V.~M. Tkachuk,
\newblock Maths-type $q$-deformed coherent states for $q>1$,
\newblock {Phys. Lett. A} \textbf{313}, 29--36 (2003).

\bibitem{Ghosh_Roy}
S.~Ghosh and P.~Roy,
\newblock ``stringy" coherent states inspired by generalized uncertainty
  principle,
\newblock {Phys. Lett. B} \textbf{711}, 423--427 (2012).

\bibitem{Ching_Ng}
C.~L. Ching and W.~K. Ng,
\newblock Generalized coherent states under deformed quantum mechanics with
  maximum momentum,
\newblock {Phys. Rev. D} \textbf{88}, 084009 (2013).

\bibitem{Pedram}
P.~Pedram,
\newblock Coherent states in gravitational quantum mechanics,
\newblock {Int. J. Mod. Phys. D} \textbf{22}, 1350004 (2013).

\bibitem{Fakhri_Hashemi}
H.~Fakhri and A.~Hashemi,
\newblock {Nonclassical properties of the $q$-coherent and $q$-cat states of
  the Biedenharn-Macfarlane $q$ oscillator with $q>1$},
\newblock {Phys. Rev. A} \textbf{93}, 013802 (2016).

\bibitem{Ramirez_Reboiro}
R.~Ram{\'\i}rez and M.~Reboiro,
\newblock Squeezed states from a quantum deformed oscillator {Hamiltonian},
\newblock {Phys. Lett. A} \textbf{380}, 1117--1124 (2016).

\bibitem{Jarvis_Lohe}
P.~D. Jarvis and M.~A. Lohe,
\newblock Quantum deformations and $q$-boson operators,
\newblock {J. Phys. A: Math. Theor.} \textbf{49}, 431001 (2016).

\bibitem{Maggiore}
M.~Maggiore,
\newblock A generalized uncertainty principle in quantum gravity,
\newblock {Phys. Lett. B} \textbf{304}, 65--69 (1993).

\bibitem{Hossenfelder_Review}
S.~Hossenfelder,
\newblock Minimal length scale scenarios for quantum gravity,
\newblock {Liv. Rev. Relat.} \textbf{16}, 2 (2013).

\bibitem{Pikovski}
I.~Pikovski et al.,
\newblock Probing Planck-scale physics with quantum optics,
\newblock {Nature Phys.} \textbf{8}, 393--397 (2012).

\bibitem{DeyNPB}
S.~Dey et al.,
\newblock Probing noncommutative theories with quantum optical experiments,
\newblock {Nucl. Phys. B} \textbf{924}, 578--587 (2017).

\bibitem{Gamboa}
J.~Gamboa, M.~Loewe and J.~C. Rojas,
\newblock Noncommutative quantum mechanics,
\newblock {Phys. Rev. D} \textbf{64}, 067901 (2001).

\bibitem{Girotti}
H.~O. Girotti,
\newblock Noncommutative quantum mechanics,
\newblock {Am. J. Phys.} \textbf{72}, 608--612 (2004).

\bibitem{Scholtz}
F.~G. Scholtz, L.~Gouba, A.~Hafver and C.~M. Rohwer,
\newblock Formulation, interpretation and application of non-commutative
  quantum mechanics,
\newblock {J. Phys. A: Math. Theor.} \textbf{42}, 175303 (2009).

\bibitem{Biedenharn}
L.~C. Biedenharn,
\newblock The quantum group {$SUq (2)$} and a $q$-analogue of the boson
  operators,
\newblock {J. Phys. A: Math. Gen.} \textbf{22}, L873 (1989).

\bibitem{Bender_Boettcher}
C.~M. Bender and S.~Boettcher,
\newblock Real spectra in {non-Hermitian Hamiltonians having
  $\mathcal{PT}$-symmetry},
\newblock {Phys. Rev. Lett.} \textbf{80}, 5243 (1998).

\bibitem{Bender_Making_Sense}
C.~M. Bender,
\newblock Making sense of {non-Hermitian Hamiltonians},
\newblock {Rep. Prog. Phys.} \textbf{70}, 947 (2007).

\bibitem{neumann}
J.~Von~Neumann and E.~Wigner,
\newblock {{\"U}ber merkw{\"u}rdige diskrete Eigenwerte. {\"U}ber das Verhalten von Eigenwerten bei adiabatischen Prozessen},
\newblock {Zhurnal Physik} \textbf{30}, 467--470 (1929).

\bibitem{friedrich1}
H.~Friedrich and D.~Wintgen,
\newblock Interfering resonances and bound states in the continuum,
\newblock {Phys. Rev. A} \textbf{32}, 3231 (1985).

\bibitem{persson}
E.~Persson, T.~Gorin and I.~Rotter,
\newblock Decay rates of resonance states at high level density,
\newblock {Phys. Rev. E} \textbf{54}, 3339 (1996).

\bibitem{wigner1}
E.~P. Wigner,
\newblock Normal form of antiunitary operators,
\newblock {J. Math. Phys.} \textbf{1}, 409 (1960).

\bibitem{bender_brody_jones}
C.~M. Bender, D.~C. Brody and H.~F. Jones,
\newblock Complex extension of quantum mechanics,
\newblock {Phys. Rev. Lett.} \textbf{89}, 270401 (2002).

\bibitem{weigert}
S.~Weigert,
\newblock $\mathcal{PT}$-symmetry and its spontaneous breakdown explained by
  anti-linearity,
\newblock {J. Phys. B: Quant. Semiclas. Opt.} \textbf{5}, S416 (2003).

\bibitem{dorey}
P.~Dorey, C.~Dunning and R.~Tateo,
\newblock A reality proof in $\mathcal{PT}$-symmetric quantum mechanics,
\newblock {Czech. J. Phys.} \textbf{54}, 35--41 (2004).

\bibitem{weigert_completeness}
S.~Weigert,
\newblock Completeness and orthonormality in $\mathcal{PT}$-symmetric quantum
  systems,
\newblock {Phys. Rev. A} \textbf{68}, 062111 (2003).

\bibitem{bender_brody_jones_prd}
C.~M. Bender, D.~C. Brody and H.~F. Jones,
\newblock Extension of $\mathcal{PT}$-symmetric quantum mechanics to quantum
  field theory with cubic interaction,
\newblock {Phys. Rev. D} \textbf{70}, 025001 (2004).

\bibitem{Bebiano}
J. da Provid{\^e}ncia, N. Bebiano and J. da Provid{\^e}ncia,
\newblock Non-Hermitian Hamiltonians with real spectrum in quantum mechanics,
\newblock {Braz. J. Phys.} \textbf{41}, 78 (2011).

\bibitem{pauli}
W.~Pauli,
\newblock {On Dirac's new method of field quantization},
\newblock {Rev. Mod. Phys.} \textbf{15}, 175 (1943).

\bibitem{sudarshan}
E.~C.~G. Sudarshan,
\newblock {Quantum mechanical systems with indefinite metric. I},
\newblock {Phys. Rev.} \textbf{123}, 2183 (1961).

\bibitem{Lee_Wick}
T.~D. Lee and G.~C. Wick,
\newblock Negative metric and the unitarity of the $s$-matrix,
\newblock {Nucl. Phys. B} \textbf{9}, 209--243 (1969).

\bibitem{dieudonne}
J.~Dieudonn{\'e},
\newblock {Quasi-Hermitian operators},
\newblock {Proc. Int. Symp. on Linear Spaces (Jerusalem, 1960), Pergamon,
  Oxford}, pp 115--122, 1961.

\bibitem{Scholtz_Geyer_Hahne}
F.~G. Scholtz, H.~B. Geyer and F.~Hahne,
\newblock {Quasi-Hermitian} operators in quantum mechanics and the variational principle,
\newblock {Ann. Phys.} \textbf{213}, 74--101 (1992).

\bibitem{mostafazadeh1}
A.~Mostafazadeh,
\newblock {Pseudo-Hermiticity versus $\mathcal{PT}$ symmetry: the necessary
  condition for the reality of the spectrum of a non-Hermitian Hamiltonian},
\newblock {J. Math. Phys.} \textbf{43}, 205 (2002).

\bibitem{dyson}
F.~J. Dyson,
\newblock General theory of spin-wave interactions,
\newblock {Phys. Rev.} \textbf{102}, 1217 (1956).

\bibitem{bagchi_quesne_roychoudhury}
B.~Bagchi, C.~Quesne and R.~Roychoudhury,
\newblock {Pseudo-Hermiticity} and some consequences of a generalized quantum
  condition,
\newblock {J. Phys. A: Math. Gen.} \textbf{38}, L647--L652 (2005).

\bibitem{znojil_geyer}
M.~Znojil and H.~B. Geyer,
\newblock Construction of a unique metric in {quasi-Hermitian} quantum
  mechanics: nonexistence of the charge operator in a 2 $\times$ 2 matrix
  model,
\newblock {Phys. Lett. B} \textbf{640}, 52--56 (2006).

\bibitem{Dey_Fring_Mathanaranjan1}
S.~Dey, A.~Fring and T.~Mathanaranjan,
\newblock {Non-Hermitian systems of Euclidean Lie algebraic type with real
  energy spectra},
\newblock {Ann. Phys.} \textbf{346}, 28--41 (2014).

\bibitem{ghatak_mandal}
A.~Ghatak and B.~P. Mandal,
\newblock Comparison of different approaches of finding the positive definite
  metric in {pseudo-Hermitian} theories,
\newblock {Commun. Theor. Phys.} \textbf{59}, 533 (2013).

\bibitem{faria_fring1}
C.~F. de~Morisson~Faria and A.~Fring,
\newblock {Isospectral Hamiltonians from Moyal products},
\newblock {Czk. J. Phys.} \textbf{56}, 899--908 (2006).

\bibitem{Guo}
A.~Guo~et al.,
\newblock Observation of {$\mathcal{PT}$}-symmetry breaking in complex optical potentials,
\newblock {Phys. Rev. Lett.} \textbf{103}, 093902 (2009).

\bibitem{ruter2010}
C.E. R{\"u}ter~et al.,
\newblock Observation of parity-time symmetry in optics,
\newblock {Nature phys.} \textbf{6}, 192 (2010).

\bibitem{regensburger}
A.~Regensburger~et al.,
\newblock Parity-time synthetic photonic lattices,
\newblock {Nature} \textbf{488}, 167 (2012).

\bibitem{feng2013}
L.~Feng~et al.,
\newblock Experimental demonstration of a unidirectional reflectionless
  parity-time metamaterial at optical frequencies,
\newblock {Nature Materials} \textbf{12}, 108 (2013).

\bibitem{lin2011}
Z.~Lin~et al.,
\newblock Unidirectional invisibility induced by $\mathcal{PT}$-symmetric periodic structures,
\newblock {Phys. Rev. Lett.} \textbf{106}, 213901 (2011).

\bibitem{chong2010}
Y.D. Chong, L.~Ge, H.~Cao and A.D. Stone,
\newblock Coherent perfect absorbers: time-reversed lasers,
\newblock {Phys. Rev. Lett.} \textbf{105}, 053901 (2010).

\bibitem{chtchelkatchev}
N.~M. Chtchelkatchev, A.~A. Golubov, T.~I. Baturina and V.~M. Vinokur,
\newblock Stimulation of the fluctuation superconductivity by {$\mathcal{PT}$} symmetry,
\newblock {Phys. Rev. Lett.} \textbf{109}, 150405 (2012).

\bibitem{bittner2012}
S.~Bittner~et al.,
\newblock {$\mathcal{PT}$} symmetry and spontaneous symmetry breaking in a
  microwave billiard,
\newblock {Phys. Rev. Lett.} \textbf{108}, 024101 (2012).

\bibitem{zheng2013}
C.~Zheng, L.~Hao and G.~L. Long,
\newblock Observation of a fast evolution in a parity-time-symmetric system,
\newblock {Phil. Trans. R. Soc. A} \textbf{371}, 20120053 (2013).

\bibitem{Moiseyev_Book}
N.~Moiseyev,
\newblock {\em {Non-Hermitian} quantum mechanics},
\newblock Cambridge Univ. Press: New York (2011).

\bibitem{Bagarello_Book}
F.~Bagarello, J.-P. Gazeau, F.~H. Szafraniec and M.~Znojil,
\newblock {\em Non-selfadjoint operators in quantum physics: {Mathematical}
  aspects},
\newblock John Wiley \& Sons: New Jersey (2015).

\bibitem{Filho_Vogel}
R.~L.~M. Filho and W.~Vogel,
\newblock Nonlinear coherent states,
\newblock {Phys. Rev. A} \textbf{54}, 4560 (1996).

\bibitem{Roy_Roy}
B.~Roy and P.~Roy,
\newblock New nonlinear coherent states and some of their nonclassical
  properties,
\newblock {J. Opt. B: Quant. Semiclas. Opt.} \textbf{2}, 65 (2000).

\bibitem{Oberhettinger}
F.~Oberhettinger,
\newblock {\em Tables of {Mellin} transforms},
\newblock Springer (2012).

\bibitem{Bergeron_Gazeau}
H.~Bergeron and J.~P. Gazeau,
\newblock Integral quantizations with two basic examples,
\newblock {Ann. Phys.} \textbf{344}, 43--68 (2014).

\bibitem{Nieto_Truax_PRL}
M.~M. Nieto and D.~R. Truax,
\newblock Squeezed states for general systems,
\newblock {Phys. Rev. Lett.} \textbf{71}, 2843 (1993).

\bibitem{Fu_Sasaki}
H-C Fu and R.~Sasaki,
\newblock Exponential and {Laguerre squeezed states for $su (1, 1)$ algebra and the Calogero-Sutherland model},
\newblock {Phys. Rev. A} \textbf{53}, 3836 (1996).

\bibitem{Angelova_Hertz_Hussin}
M.~Angelova, A.~Hertz and V.~Hussin,
\newblock Squeezed coherent states and the one-dimensional {Morse} quantum
  system,
\newblock {J. Phys. A: Math. Theor.} \textbf{45}, 244007 (2012).

\bibitem{Sivakumar_PACS}
S.~Sivakumar,
\newblock Photon-added coherent states as nonlinear coherent states,
\newblock {J. Phys. A: Math. Gen.} \textbf{32}, 3441 (1999).

\bibitem{Duc_Noh}
T.~M. Duc and J.~Noh,
\newblock Higher-order properties of photon-added coherent states,
\newblock {Opt. Commun.} \textbf{281}, 2842--2848 (2008).

\bibitem{Zavatta_Viciani_Bellini}
A.~Zavatta, S.~Viciani and M.~Bellini,
\newblock Quantum-to-classical transition with single-photon-added coherent
  states of light,
\newblock {Science} \textbf{306}, 660--662 (2004).

\bibitem{Hillery}
M.~Hillery,
\newblock Amplitude-squared squeezing of the electromagnetic field,
\newblock {Phys. Rev. A} \textbf{36}, 3796 (1987).

\bibitem{Hong_Mandel}
C.~K. Hong and L.~Mandel,
\newblock Higher-order squeezing of a quantum field,
\newblock {Phys. Rev. Lett.} \textbf{54}, 323 (1985).

\bibitem{Graefe_PRL}
E.~M. Graefe, H.~J. Korsch and A.~E. Niederle,
\newblock Mean-field dynamics of a {non-Hermitian Bose-Hubbard} dimer,
\newblock {Phys. Rev. Lett.} \textbf{101}, 150408 (2008).

\bibitem{Jayakrishnan_Dey}
M.~P. Jayakrishnan, S.~Dey, M.~Faizal and C.~Sudheesh,
\newblock $q$-deformed quadrature operator and optical tomogram,
\newblock {Ann. Phys.} \textbf{385}, 584--590 (2017).

\bibitem{Graefe_Schubert}
E.-M. Graefe and R.~Schubert,
\newblock Complexified coherent states and quantum evolution with
  {non-Hermitian Hamiltonians},
\newblock {J. Phys. A: Math. Theor.} \textbf{45}, 244033 (2012).

\bibitem{Cherbal}
O.~Cherbal, M.~Drir, M.~Maamache and D.~A. Trifonov,
\newblock Fermionic coherent states for {pseudo-Hermitian two-level systems},
\newblock {J. Phys. A: Math. Theor.} \textbf{40}, 1835 (2007).

\bibitem{Beckers}
J.~Beckers, N.~Debergh, J.~F. Cari{\~n}ena and G.~Marmo,
\newblock {Non-Hermitian oscillator-like Hamiltonians and $\lambda$-coherent
  states revisited},
\newblock {Mod. Phys. Lett. A} \textbf{16}, 91--98 (2001).

\bibitem{Kandirmaz}
N.~Kandirmaz and R.~Sever,
\newblock Coherent states for {PT-/non-PT-symmetric and non-Hermitian Morse
  potentials via the path integral method},
\newblock {Phys. Scr.} \textbf{81}, 035302 (2010).

\bibitem{Bagarello_BiCoherent}
F. Bagarello,
\newblock Intertwining operators for non-self-adjoint Hamiltonians and bicoherent states,
\newblock {J. Math. Phys.} \textbf{57}, 103501 (2016).

\bibitem{Zelaya2017}
K.~Zelaya, S.~Dey, V.~Hussin and O.~Rosas-Ortiz,
\newblock Nonclassical states for {non-Hermitian} hamiltonians with the
  oscillator spectrum,
\newblock {arXiv:1707.05367}.

\bibitem{Yahiaoui}
S.-A. Yahiaoui and M.~Bentaiba,
\newblock New position-dependent effective mass coherent states for a
  generalized shifted harmonic oscillator,
\newblock {J. Phys. A: Math. Theor.} \textbf{47}, 025301 (2013).

\bibitem{Wang_Goorskey_Xiao}
H.~Wang, D.~Goorskey and M.~Xiao,
\newblock Enhanced {Kerr} nonlinearity via atomic coherence in a three-level
  atomic system,
\newblock {Phys. Rev. Lett.} \textbf{87}, 073601 (2001).

\bibitem{Gambetta2006}
A.~Gambetta~et al.,
\newblock Real-time observation of nonlinear coherent phonon dynamics in
  single-walled carbon nanotubes,
\newblock {Nature Phys.} \textbf{2}, 515 (2006).

\bibitem{Yan_Zhu_Li}
Y.~Yan, J.-P. Zhu and G.-X. Li,
\newblock Preparation of a nonlinear coherent state of the mechanical resonator in an optomechanical microcavity,
\newblock {Opt. Exp.} \textbf{24}, 13590--13609 (2016).

\end{thebibliography}


\end{document}